\documentclass[aps,prl,twocolumn,10pt,amsmath,amssymb,nofootinbib,superscriptaddress,floatfix]{revtex4-2}

\usepackage{graphicx}
\usepackage{dcolumn}
\usepackage{bm}
\usepackage{dsfont}
\usepackage{bbm}
\usepackage{textgreek}
\usepackage{mathtools}
\usepackage{placeins}
\usepackage{xcolor}
\usepackage{capt-of}
\usepackage{tikz-cd}
\definecolor{figblue}{HTML}{1F77B4}
\definecolor{figorange}{HTML}{FF7F0E}
\definecolor{figgreen}{HTML}{1F9D55}
\definecolor{figred}{HTML}{D62728}

\newcommand{\figbar}[1]{%
  \protect\raisebox{0.25ex}{\textcolor{#1}{\rule{1.4em}{0.45ex}}}%
}
\newcommand{\figdot}[1]{%
  \protect\textcolor{#1}{\ensuremath{\bullet}}%
}

\newcommand{\ben}{\begin{equation}}
\newcommand{\een}{\end{equation}}
\newcommand{\be}{\begin{equation}}
\newcommand{\ee}{\end{equation}}
\newcommand{\bea}{\begin{eqnarray}}
\newcommand{\eea}{\end{eqnarray}}
\newcommand{\ba}{\begin{eqnarray}}
\newcommand{\ea}{\end{eqnarray}}

\newcommand{\beq}{\begin{equation}}
\newcommand{\eeq}{\end{equation}}
\newcommand{\beqa}{\begin{eqnarray}}
\newcommand{\eeqa}{\end{eqnarray}}
\newcommand{\beqar}{\begin{eqnarray*}}
\newcommand{\eeqar}{\end{eqnarray*}}

\def\t6 {T_\mt{D6}}

\newcommand{\mt}[1]{\textrm{\tiny #1}}

\def\cale         {{\cal E}}

\def\ee           {{\rm e}}

\def\sqr#1#2{{\vcenter{\vbox{\hrule height.#2pt
 \hbox{\vrule width.#2pt height#1pt \kern#1pt
 \vrule width.#2pt}\hrule height.#2pt}}}}

\definecolor{chartedge}{HTML}{3E6B89}
\definecolor{chartfill}{HTML}{EAF3F8}

\def\ee{\cale}

\def\aa1{\phi}
\def\cc1{\psi}

\def\ben{\begin{equation}}
\def\een{\end{equation}}
\def\bea{\begin{eqnarray}}
\def\eea{\end{eqnarray}}

\usepackage{silence}
\usepackage[bookmarks=false]{hyperref}
\hypersetup{pdfstartview=FitH,pdfhighlight=/O,colorlinks=true, allcolors=blue}
\usepackage{orcidlink}

\begin{document}

\title{Fixed Points, Floquet Entanglement Asymmetry, and Quantum Mpemba Effects}

\author{Jayashish Das${}^{\orcidlink{0009-0009-8199-8901}}$}
\email{jayashish.das[at]saha.ac.in}
\affiliation{Theory Division, Saha Institute of Nuclear Physics,
A CI of Homi Bhabha National Institute, 1/AF Bidhannagar,
Kolkata 700064, India}

\author{Filiberto Ares${}^{\orcidlink{0000-0001-5062-2332
}}$}
\email{faresase[at]sissa.it}
\affiliation{SISSA and INFN Sezione di Trieste,
via Bonomea 265, 34136 Trieste, Italy}

\author{Arnab Kundu${}^{\orcidlink{0000-0002-1994-3346}}$}
\email{arnab.kundu[at]saha.ac.in}
\affiliation{Theory Division, Saha Institute of Nuclear Physics,
A CI of Homi Bhabha National Institute, 1/AF Bidhannagar,
Kolkata 700064, India}


\begin{abstract}

We investigate the dynamics of entanglement asymmetry in periodically driven two-dimensional conformal field theories with a global U$(1)$ symmetry, using a dynamical-system description based on iterated conformal maps, in particular Möbius maps. Starting from a symmetry-breaking excited state prepared by a local operator insertion, we show that a broad range of nonequilibrium phenomena, including local symmetry restoration, the quantum Mpemba effect, and its inverse, admit a simple geometric description in terms of the invariant data of the conformal map. We explicitly demonstrate this using Möbius maps, for which the invariant structure is described by the conjugacy classes. In particular, the dynamics of entanglement asymmetry is determined by the relative positions of the state-preparing operator insertion, the subsystem, and the fixed points, yielding distinct patterns of growth, decay, oscillation, and saturation. We further argue that this geometric picture extends to general holomorphic maps, establishing the conformal-map dynamics as a natural framework for organizing the phenomenology of entanglement asymmetry in periodically driven two-dimensional conformal field theories.
  
\end{abstract}

\maketitle

{\it Introduction:} Periodically driven systems provide us with a remarkably broad class of novel non-equilibrium phenomena, which have no equilibrium analogs. These include heating, dynamical phase transitions, localization; see {\it e.g.}~\cite{rev2,rev3,rev8,rev9}, among others. However, a general organizing principle that predicts the qualitative features of these phenomena, irrespective of microscopic details, remains elusive. It is therefore highly desirable to distill invariant structures that can classify the resulting dynamical behavior, rather than exploring individual models separately. 

Conformal Field Theories (CFTs), especially in two spacetime dimensions, constitute a large class of systems in which controlled (semi-) analytical calculations of correlation functions can be carried out. This framework emerges naturally at continuous phase transition points and is now experimentally realizable on quantum simulators~\cite{joshi23, sun2026cftexp, mo2026observingconformalfloquetdynamics}. For this class of systems, a periodic drive can be implemented as a sequence of Möbius maps, and more generally of holomorphic maps. The corresponding stroboscopic dynamics is then described by the iteration of elements of the PSL$(2,{\mathbb R})$ group, whose conjugacy classes distinguish a heating phase, a non-heating phase, and a phase-transition line~\cite{2dpapers1}. This operator-algebraic framework in CFT already hints at a universal description of non-equilibrium driven dynamics, see {\it e.g.}~\cite{dynph1, dynph2, 2dpapers2,2dpapers3,2dpapers4,2dpapers5,2dpapers6, 2dpapers7, 2dpapers9,2dpapers10,2dpapers11,2dpapers12, Das:2022jrr} for several interesting applications of this framework in two dimensions, \cite{2dcurved1,2dcurved2,2dcurved3, holo1,holo2,holo3,holo4,holo5} for CFTs with holographic duals, \cite{hd1, hd2} for CFTs in higher dimensions, to name a few. In this sense, Floquet evolution can be recast as a problem in conformal-map dynamics.

 Such sequences of conformal maps contain invariant data. Among these, the fixed points and their stability determine the asymptotic dynamics of the corresponding map. For Möbius maps, the local data near the fixed points in fact characterize the global dynamics, since such maps have no critical points. This suggests the existence of a geometric classification.

In this work, we focus on a quantum information measure, namely the entanglement asymmetry; see {\it e.g.}~\cite{ares23asymmetry, bartlett07, vaccaro08, gour09, marvian14}. Entanglement asymmetry provides a quantitative measure of the restoration or breaking of an internal symmetry during non-equilibrium evolution. While its dynamics has been investigated in a variety of nonequilibrium settings, including experiments~\cite{joshi24, xu2026EAexp, Yang26Pageexp}, its behavior under Floquet driving in a CFT remains largely unexplored; see, however, Ref.~\cite{Banerjee:2024zqb} for a study of entanglement asymmetry starting from a boundary state with global symmetry breaking. Here, we investigate entanglement asymmetry in a Floquet CFT starting from an initial state prepared by a local operator insertion that breaks the symmetry~\cite{Benini:2024xjv}. Studies of quantum quenches have revealed a rich variety of dynamical behaviors for the entanglement asymmetry (see {\it e.g.}~\cite{Benini:2024xjv, rylands24, Murciano2024, Yamashika_2024,  Ares_2025, liu24, turkeshi25, ares25circ, castroalvaredo26, yamashika26, russotto26, summer26, muller26, aditya26, hammer26, vescovo26}), but their origin has so far been understood primarily in terms of the underlying quench dynamics. We show that, within the Floquet CFT setting, the resulting dynamics can instead be traced to the invariant geometry of the conformal dynamical map and is classified by its conjugacy class, rather than by the detailed microscopic form of the Hamiltonian. As examples of this general principle, we demonstrate that this geometric structure governs several non-equilibrium phenomena, including local symmetry restoration and its absence~\cite{Ares:2023kcz, Caceffo_2024, yamashika25, hara26, digiulio25}, the emergence of the quantum Mpemba effect and its inverse analogue~\cite{ares25nat, teza26, yu2025, calabrese26}, as well as multiple crossings of the entanglement asymmetry between initially distinct states~\cite{Chalas_2024, chatterjee24, mcroberts26}.
 
A particular CFT describes a specific universality class in the context of a continuous phase transition. This gives two complementary perspectives on our results. First, within a given universality class, the dynamics of entanglement asymmetry is characterized by the fixed-point structure of the conformal dynamical map. For an experimentally realizable driving protocol, the fixed-point structure therefore determines the dynamics of entanglement asymmetry throughout the corresponding universality class. Second, several phenomena of entanglement asymmetry observed in other non-equilibrium settings also arise within the driven CFT framework, where they admit a unified geometric description in terms of Möbius, and more generally holomorphic, maps. The resulting classical dynamical system provides a simple geometric organization of the possible dynamical behaviors of the entanglement asymmetry.

{\it CFT \& Driving}: Let us begin with a two-dimensional CFT defined on the Euclidean cylinder ${\mathbb R} \times S^1$. The cylinder coordinate is $w = \tau + i x$, with $x \sim x + L$, where $L$ is the circumference of $S^1$. Using the exponential map, $u=e^{2\pi w/L}$, the cylinder is mapped to the complex plane. Consider a sequence of unitary evolutions on the cylinder $U_i =e^{-i H_i T_i}$, where $H_i$ is the Hamiltonian governing the $i$-th step and $T_i$ is the corresponding evolution time. We consider a family of inhomogeneous Hamiltonians:
\begin{eqnarray}\label{eq:inh_ham_CFT}
    H_i = \int_0^L  dx \, v_i(x) T_{00} (x) \ , 
\end{eqnarray}
where $v_i(x)$ is a smooth periodic deformation and $T_{00}=(T + {\bar T})/(2\pi)$ is the energy density of the CFT stress-energy tensor. 
For the \(SL(2,\mathbb{R})\) subclass considered below, each driving step induces a Möbius transformation on the complex plane,~\cite{2dpapers1}:
\begin{eqnarray}
    u \mapsto f_i(u) = \frac{a_i u + b_i}{c_i u + d_i} \ , \label{mobius}
\end{eqnarray}
where $\{a_i, b_i, c_i, d_i\}$ are determined by $H_i$ and $T_i$. The sequence of unitary evolutions therefore corresponds to the composition of the associated conformal maps. Denoting the map corresponding to one complete driving cycle by $f$, the position of an initial point $u_0$ after $N$ cycles is:
\begin{eqnarray}\label{eq:it_moebius}
    u_N = f^N(u_0)=\underbracket{f\left(f\left(\ldots f\left(u_0 \right) \right) \right)}_{N-{\rm times}} \ .
\end{eqnarray}
The corresponding evolution of a primary operator on the cylinder with conformal dimensions $(h, \bar{h})$ can then be written in the Heisenberg picture as:
\begin{equation}
U_N^{\dagger}\,O(w,\bar w)\,U_N =
  \left(\frac{\partial w_N}{\partial w}\right)^{\!h}
  \left(\frac{\partial \bar w_N}{\partial \bar w}\right)^{\!\bar h}
  O\big(w_N,\bar w_N\big)\,,
  \label{eq:primary-time-evolution}
\end{equation}
where $U_N$ denotes the evolution over $N$ cycles and $w_N$ is defined by $u_N=e^{2\pi w_N/L}$. 

Before discussing the specific observable, let us summarize the fixed-point structure of the map (\ref{mobius}) which also governs the localization of energy and entanglement in the heating phase~\cite{Fan_2020}. The fixed point, $u_*$, and the multiplier, $\lambda$, are defined as:
\begin{eqnarray}
    f(u_*) = u_* \ , \quad \lambda = \left. \frac{df}{du} \right|_{u_*} \ . 
\end{eqnarray}
The fixed points of the M\"{o}bius map can be classified into three different conjugacy classes. (i) Elliptic: one fixed point is inside the Poincar\'e disk and the other one is outside, {\it i.e.}~they are related by inversion across the unit circle,. In this case, $|\lambda|=1$ and therefore near the fixed point the dynamics neither converges nor diverges. This typically results in persistent oscillatory behavior. (ii) Hyperbolic: there are two distinct fixed points on the boundary of the Poincar\'{e} disk. For one of them $|\lambda| <1$ and for the other one $|\lambda|>1$. The former is an attractive fixed point, and the latter is a repulsive fixed point. (iii) Parabolic: Two distinct fixed points on the Poincaré disk coalesce into a double fixed point, with $\lambda=1$. While, for the elliptic class, $f(u)-u_* \sim e^{i\theta} (u-u_*)$, so the distance from the fixed point remains constant, for the parabolic case, $u_N - u_* \sim N^{-1}$. Hence, the parabolic fixed point is marginally attractive.

{\it Our Setup}: In this letter, we consider a two-dimensional CFT with a global U$(1)$ symmetry. Given a density matrix $\rho$ of the full system and a bipartition into a spatial region $A$ and its complement $\bar{A}$ at a constant Euclidean time slice $\tau=0$ on the cylinder, we define the reduced density matrix $\rho_A = {\rm Tr}_{\bar{A}}(\rho)$.  We initialize the system in a state that breaks the U$(1)$ symmetry and investigate how this symmetry breaking evolves in the subsystem $A$ under a Floquet evolution that preserves the symmetry. Despite the globally unitary evolution, the symmetry can be dynamically restored in the subsystem $A$. The symmetry of the reduced state is characterized by the commutator $[\rho_A, Q_A]$, where $Q_A$ is the conserved U$(1)$ charge restricted to $A$. This symmetry breaking, and its possible restoration, is quantified by the entanglement asymmetry, defined as the relative entropy:
\begin{equation}\label{eq:def_EA}
    \Delta S_A = S_A \left(\rho_A || \rho_{A,Q} \right)={\rm Tr}(\rho_A(\log\rho_A-\log\rho_{A, Q})) \ ,
\end{equation}
where $\rho_{A, Q}$ is the symmetrization of $\rho_A$ and, by construction, satisfies $[\rho_{A, Q}, Q_A]=0$. In particular, for a U$(1)$ global symmetry, $\rho_{A, Q}$ is obtained by averaging over a U$(1)$ orbit. The entanglement asymmetry vanishes if and only if $\rho_A$ is symmetric, {\it i.e}, $[\rho_A, Q_A]=0$.
In our setting, the entanglement asymmetry is given by a functional of the iterated M\"{o}bius maps~\eqref{eq:it_moebius}. 

To evaluate the entanglement asymmetry, it is useful to apply the replica trick. Eq.~\eqref{eq:def_EA} can be recovered in the limit $n\to 1$ from the difference between the R\'enyi-$n$ entanglement entropies of $\rho_{A, Q}$ and $\rho_A$~\cite{ares23asymmetry}. The R\'enyi entropy of $\rho_A$ corresponds to the CFT path integral on an $n$-sheeted Riemann surface with a branch cut along $A$~\cite{cc04, cc09}. For $\rho_{A, Q}$, the same geometry is used, with the U$(1)$ group elements over which we average inserted as defects along $A$~\cite{fossati24, chenchen24, kusuki25, fossati25, lastres25, venuti25, valdecasas25,fossati26}. 
In general, it is difficult to obtain generic results for this. However, for a specific class of excited states obtained from symmetry-breaking coherent excitations of the vacuum, the calculation simplifies substantially. Such states are generated by the insertion of a local operator $O(w) = e^{i \kappa V(w)}$, where $V$ is a charged spinless primary operator of scaling dimension $\Delta=h+\bar{h}$, the insertion point has coordinates $w=-\tau_0+i\eta$, and $\kappa$ is a free parameter~\cite{Benini:2024xjv}. This parameter allows for a controlled perturbative expansion and, at leading order in $\kappa\ll 1$, $\Delta S_A$ is determined by a set of two-point correlation functions of the operator $V$. The sequence of Möbius maps dynamically evolves the insertion point of the local operator $O$ with respect to the endpoints of the subsystem $A$, which remain fixed along the evolution. Thus, the relevant dynamical data are: (i) the endpoints of $A$, (ii) the insertion point of $O$, and (iii) the fixed points of the Möbius map, which are determined by its conjugacy class. While (i) and (ii) can be chosen freely, the subsequent evolution of (ii) and the fixed-point structure in (iii) are determined by the Hamiltonian evolution.

{\it Phenomenology of Entanglement Asymmetry}: Let us now demonstrate the phenomenology of entanglement asymmetry with a concrete example. We consider the two-step driving protocol $U_N=(e^{-iT_1H_1} e^{-i T_0H_0})^N$ where $H_0=H(\theta=0)$ and $H_1=H(\theta\neq 0)$, and $H(\theta)$ is the ${\rm SL}_2$ deformed Hamiltonian:
\begin{equation}
H(\theta)=\frac{2\pi}{L}\Big[L_0-\tanh(2\theta)\,
     \frac{L_1+L_{-1}}{2}-\frac{c}{24}\Big]+{\rm h.c.},
\end{equation}
with $\theta>0$ and $L_k$ the Virasoro generators.
This corresponds to $v(x)=1-\tanh(2\theta)\cos(2\pi x/L)$ in Eq.~\eqref{eq:inh_ham_CFT}. 

The fixed points of the corresponding Möbius map depend on its conjugacy class. For the hyperbolic class, the attractive and repulsive fixed points are located at $x_{\rm att}=L/2$ and $x_{\rm rep}=0$, respectively. In the large-$N$ stroboscopic limit, the resulting entanglement asymmetry behaves as:
\begin{equation}
    \Delta S_{A}(N)
    \sim
    \begin{cases}
        e^{-8\Delta N\theta},
        & x_{\rm att}\in\overline A,
        \\[5pt]
        e^{4N\theta},
        & x_{\rm att}\in A,
        \\[5pt]
        \mathrm{constant},
        & \{x_{\rm att},x_{\rm rep}\}=\partial A,
        \\[5pt]
        \Delta S_{A,\infty}-c_+e^{-4N\theta},
        & x_{\rm att}\in\partial A,
                    x_{\rm rep}\in\overline A,
        \\[5pt]
        \Delta S_{A,\infty}+c_-e^{-4N\theta},
        & x_{\rm att}\in\partial A,
                    x_{\rm rep}\in A,
        \end{cases} \label{Scases}
\end{equation}
when truncating at the leading-order contribution in $\kappa\ll 1$. Here, $\Delta S_{A,\infty}$ and $c_{\pm}$ are constants whose details are not important for this discussion. 
This result captures the main message of this letter: the large-time behavior of entanglement asymmetry is determined by the kinematic arrangement of the fixed points and the endpoints of the subsystem. 
The different asymptotic behaviors have a direct geometric interpretation. When the attractive fixed point lies inside $A$, the operator-insertion is driven towards it and the entanglement asymmetry grows exponentially, implying that the symmetry is not restored in the subsystem. Conversely, when the attractive fixed point lies in $\bar A$, the entanglement asymmetry decays exponentially to zero, signaling the dynamical restoration of the symmetry in $A$. Finally, when the attractive fixed point coincides with an end-point of $A$, the entanglement asymmetry approaches a finite value, and the symmetry is not restored.

\begin{figure*}[!t]
    \centering

    \includegraphics[width=\textwidth]{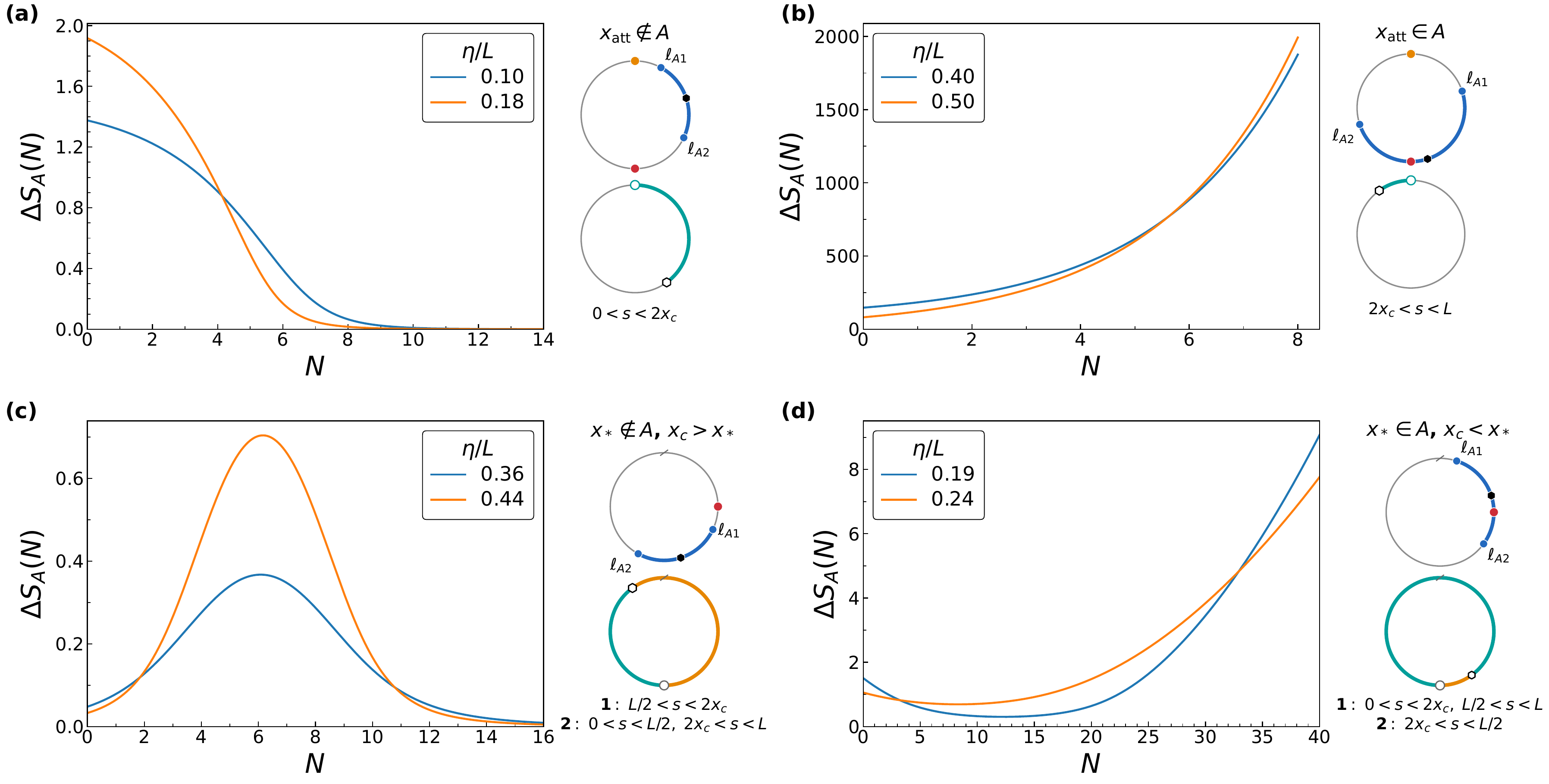}

    \caption{
   Evolution of the entanglement asymmetry $\Delta S_A(N)$ with the number of drive cycles $N$ for two initial insertion points $\eta$ in the heating phase (upper panels) and at the phase transition (lower panels). We set $\Delta=1$ and $\theta=0.1$ throughout, with $\tau_0/L=(0.1,\,0.002,\,0.16,\,0.07)$, in panels (a)--(d), respectively. The driving periods are chosen as $\left(T_0/L,T_1/L\right) =\left(1/2,\cosh(2\theta)/2\right)$ for heating phase in panels (a) and (b), and $\left(T_0/L,T_1/L\right) =\left(1/2,\cosh(2\theta)\arcsin[1/\cosh(2\theta)]/\pi\right)$ for phase transition in panels (c) and (d). (a) When the attractive fixed point lies outside the subsystem, the asymmetry decays to zero. For the values of $\eta$ chosen, the initially more asymmetric state relaxes faster, realizing the quantum Mpemba effect. (b) When the attractive fixed point lies inside the subsystem, the asymmetry grows exponentially. In this particular case, the initially more symmetric state eventually becomes more asymmetric, realizing the inverse Mpemba effect. At the phase transition, the neutral fixed point lies outside the subsystem in panel~(c) and inside in panel~(d). In both cases, the nonmonotonic evolution of the asymmetry yields two crossings. In the diagrams next to each plot, \figbar{figblue} represents the subsystem $A$, with \figdot{figblue} marking its endpoints $\ell_{A1}$ and $\ell_{A2}$. The arc \figbar{figgreen} denotes the range of $s=\eta_1+\eta_2$ for which one crossing occurs, whereas \figbar{figorange} is the range for which two crossings occur. In panels (a)-(b), the symbols \figdot{figred} and \figdot{figorange} represent $x_{\rm att}$ and $x_{\rm rep}$, respectively; in panels (c)-(d), \figdot{figred} represents the neutral fixed point $x_\ast$. The filled and open black symbols denote the subsystem center $x_c$ and $2x_c$, respectively, while the open green symbols represent the remaining boundaries separating ranges of $s$ yielding zero, one, or two crossings.}
    \label{fig:mpemba_crossings}
\end{figure*}
A similar picture emerges in the parabolic class. In the large stroboscopic time limit, the entanglement asymmetry exhibits the following universal behavior when $\kappa\ll 1$:
\begin{equation}
    \Delta S_{A}(N)
    \sim
    \begin{cases}
        N^{-4\Delta},
        & x_*\notin A,
        \\[2pt]
        N^{-2\Delta},
        & x_*=\ell_{A_2},
        \\[2pt]
        N,
        & x_*=\ell_{A_1},
        \\[2pt]
        N^2,
        & x_*\in A ,
    \end{cases}
    \label{eq:PT-complete-classification}
\end{equation}
where $x_*=L/4$ denotes the location of the parabolic fixed point (weakly attracting), while $0<\ell_{A_{1}}<\ell_{A_{2}}<L$ are the endpoints of the subsystem $A$. Intuitively, the operator insertion is driven towards the attractive fixed point. When the subsystem contains the fixed point, the corresponding asymmetry therefore diverges, whereas a fixed point outside the subsystem leads to symmetry restoration. For the cases discussed above, these growth and decay behaviors are exponential or power-law, depending on the conjugacy class. In the elliptic class, the corresponding dynamics is instead oscillatory. 

So far, we have focused on the asymptotic stroboscopic time behavior. We can also obtain the full time evolution of the entanglement asymmetry in the perturbative regime $\kappa\ll 1$. The perturbative parameter is fixed as $\kappa=\epsilon\,\ell_{\tau_0}^{\Delta}$, where
$\ell_{\tau_0}=\frac{L}{2\pi}\sinh\left(2\pi\tau_0/L\right)$ the cylinder length associated with the Euclidean separation $2\tau_0$ and $\epsilon\ll1$ is dimensionless; we plot the asymmetry $\Delta S_A(N)$ divided by $\epsilon^2$. In Fig.~\ref{fig:mpemba_crossings}, we show its evolution for the hyperbolic (upper panels) and parabolic (lower panels) classes, for two different initial positions of the operator insertion. As discussed above, within each conjugacy class, the restoration of the symmetry is determined by the relative position of the Möbius fixed points and the subsystem. The full time evolution reveals additional phenomena. In Fig.~\ref{fig:mpemba_crossings}(a), the state with the initially larger asymmetry restores the symmetry faster, signaling the occurrence of the quantum Mpemba effect. Panel~(b) shows the inverse effect, where the initially more symmetric state eventually develops a larger asymmetry than the initially more asymmetric state, analogous to inverse Mpemba effects~\cite{lu17, kumar22, AharonyShapira:2024nrt,Sugimoto:2025pgk}. A single crossing between the asymmetries of two initially distinct states provides a characteristic signature of the Mpemba effect. In our setting, the occurrence of such a crossing can be characterized by a simple analytic criterion.

To state this criterion, let $\eta_{1,2}$ be the spatial points in which the operator $V$ is inserted for the two initial states, and introduce the scale $x_c = (\ell_{A_1}+ \ell_{A_2})/2$, corresponding to the midpoint of the subsystem $A$. For example, in the hyperbolic phase, the condition for a single crossing is:
\begin{eqnarray}
    \begin{aligned}
        x_{\rm att}\notin A:
        &\qquad
        0<\eta_1+\eta_2<2x_c \ ,
        \\[2pt]
        x_{\rm att}\in A:
        &\qquad
        2x_c<\eta_1+\eta_2<L \ ,
        \\[2pt]
        x_{\rm att}\in\partial A \ ,\quad
        x_{\rm rep}\notin A:
        &\qquad
        \frac{L}{2}<\eta_1+\eta_2<2x_c \ ,
        \\[2pt]
        x_{\rm att}\in\partial A \ ,\quad
        x_{\rm rep}\in A:
        &\qquad
        2x_c<\eta_1+\eta_2<\frac{L}{2} \ .
    \end{aligned}
    \label{eq:eta-crossing-summary}
\end{eqnarray}
A similar condition can be obtained in the parabolic class, although the resulting expression is algebraically more involved; see the companion paper \cite{Das:toappear} for details. Multiple crossings can also occur in both the hyperbolic and parabolic classes, as illustrated by the numerical results in Fig.~\ref{fig:mpemba_crossings}. For Möbius maps, at most two crossings can occur. In the elliptic case, by contrast, infinitely many crossings can occur, reflecting the absence of a stationary value of the asymmetry. In summary, the emergence of the quantum Mpemba effect reduces to a point-ordering problem on the circle, as shown schematically in Fig.~\ref{fig:mpemba_crossings}.

The geometric picture described above admits a natural generalization. For the purpose of classifying the spacetime motion of operator insertions and its fixed-point structure, it is sufficient to restrict the conformal evolution generated by a spatially inhomogeneous Hamiltonian~\eqref{eq:inh_ham_CFT} with a smooth real deformation profile to the $\tau=0$ slice. Under the cylinder-to-plane map, the $\tau=0$ slice is mapped to the unit circle, and the restriction of the finite-time conformal evolution defines a smooth orientation-preserving diffeomorphism, $f\in{\rm Diff}(S^1)$, which necessarily has degree one. If all its fixed points are isolated and hyperbolic, the fixed-point index theorem gives \(n_{\rm att}-n_{\rm rep}= 1-\deg(f)=0,\) so that the attractive and repulsive fixed points occur in equal numbers and alternate around the circle. Apart from these constraints, a general circle diffeomorphism may possess an arbitrary finite number of attractive-repulsive fixed-point pairs. The long-time behavior of the asymmetry is therefore expected to be qualitatively similar to Eq.~\eqref{Scases}, as long as attractive and repulsive fixed points exist. A similar argument applies to neutral fixed points.

Interestingly, the finite stroboscopic time dynamics, including the quantum Mpemba effect, is also constrained by the map and its fixed-point structure. In particular, multiple crossings between the asymmetry curves can be realized by appropriately chosen maps, generalizing the conditions in Eq.~\eqref{eq:eta-crossing-summary}. Thus, nontrivial dynamical effects in symmetry restoration for different initial states can be recast in terms of the map and its fixed points. What is even richer is that, beyond M\"obius maps, the fixed-point data and their corresponding multipliers do not uniquely determine the map. Hence, while the possible types of fixed points are therefore limited, the resulting phenomenology of entanglement asymmetry can be considerably larger.

{\it Conclusions \& Outlook}: We have shown that Floquet protocols generating the same Möbius, or more generally holomorphic, map exhibit the same qualitative, universal dynamics of the entanglement asymmetry. The asymptotic behavior is governed by the fixed points and their multipliers: for a given subsystem, they determine whether the entanglement asymmetry grows, decays, oscillates, or saturates. For general conformal maps, there can be multiple attractive fixed points. Depending on the operator insertion that prepares the initial state and the choice of subsystem, different fixed points can govern the dynamics of the asymmetry, resulting in competing symmetry-restoration channels. This opens the possibility of richer phenomena, including Mpemba effects with multiple crossings. 

This space of possibilities also includes, {\it e.g.}, periodic maps. An attracting cycle can produce an asymmetry pattern with the period of the cycle, generalizing the oscillatory behavior of the elliptic Möbius class. Similarly, neutral fixed points, generalizing the parabolic class of the Möbius maps, are expected to be associated with slow relaxation. More generally, it would be interesting to investigate the effects of quasi-periodic or chaotic invariant sets on the dynamics of entanglement asymmetry. 

It is also relevant to ask whether the higher Rényi asymmetries, which are experimentally accessible~~\cite{joshi24, xu2026EAexp, Yang26Pageexp}, exhibit similar physics. Although the corresponding replica correlators give rise to expressions that are difficult to evaluate explicitly, our general arguments suggest that their qualitative behavior should still be controlled by the invariant data of the underlying conformal map. Rényi asymmetries would thus provide a probe of this invariant data beyond the von Neumann limit. Testing this picture at general Rényi index, for multiple local symmetry-breaking operator insertions, and in explicit lattice realizations of inhomogeneously driven critical systems are natural directions for future work.

\begin{sloppypar}
\noindent {\bf Acknowledgements:} The authors thank Diptarka Das, Sumit R. Das, Krishnendu Sengupta for useful
discussions on related topics. A.K. further acknowledges the support of the Humboldt Research Fellowship for Experienced Researchers by the Alexander von Humboldt Foundation and for the hospitality of Theoretical Physics III, Department
of Physics and Astronomy, Julius-Maximilians-Universit\"{a}t W\"{u}rzburg and the support from
the ICTP through the Associates Programme (2024-2030) during the course of this work. F.A. acknowledges support from the European Research Council
under the Advanced Grant no.~101199196 (MOSE). The draft preparation made use of AI coding and assistance, including OpenAI Codex and ChatGPT, under human supervision. All physics assumptions and conclusions are the responsibility of the authors. 
\end{sloppypar}

\bibliography{biblio}

\begin{thebibliography}{84}%
\makeatletter
\providecommand \@ifxundefined [1]{%
 \@ifx{#1\undefined}
}%
\providecommand \@ifnum [1]{%
 \ifnum #1\expandafter \@firstoftwo
 \else \expandafter \@secondoftwo
 \fi
}%
\providecommand \@ifx [1]{%
 \ifx #1\expandafter \@firstoftwo
 \else \expandafter \@secondoftwo
 \fi
}%
\providecommand \natexlab [1]{#1}%
\providecommand \enquote  [1]{``#1''}%
\providecommand \bibnamefont  [1]{#1}%
\providecommand \bibfnamefont [1]{#1}%
\providecommand \citenamefont [1]{#1}%
\providecommand \href@noop [0]{\@secondoftwo}%
\providecommand \href [0]{\begingroup \@sanitize@url \@href}%
\providecommand \@href[1]{\@@startlink{#1}\@@href}%
\providecommand \@@href[1]{\endgroup#1\@@endlink}%
\providecommand \@sanitize@url [0]{\catcode `\\12\catcode `\$12\catcode `\&12\catcode `\#12\catcode `\^12\catcode `\_12\catcode `\%12\relax}%
\providecommand \@@startlink[1]{}%
\providecommand \@@endlink[0]{}%
\providecommand \url  [0]{\begingroup\@sanitize@url \@url }%
\providecommand \@url [1]{\endgroup\@href {#1}{\urlprefix }}%
\providecommand \urlprefix  [0]{URL }%
\providecommand \Eprint [0]{\href }%
\providecommand \doibase [0]{https://doi.org/}%
\providecommand \selectlanguage [0]{\@gobble}%
\providecommand \bibinfo  [0]{\@secondoftwo}%
\providecommand \bibfield  [0]{\@secondoftwo}%
\providecommand \translation [1]{[#1]}%
\providecommand \BibitemOpen [0]{}%
\providecommand \bibitemStop [0]{}%
\providecommand \bibitemNoStop [0]{.\EOS\space}%
\providecommand \EOS [0]{\spacefactor3000\relax}%
\providecommand \BibitemShut  [1]{\csname bibitem#1\endcsname}%
\let\auto@bib@innerbib\@empty
\bibitem [{\citenamefont {Polkovnikov}\ \emph {et~al.}(2011)\citenamefont {Polkovnikov}, \citenamefont {Sengupta}, \citenamefont {Silva},\ and\ \citenamefont {Vengalattore}}]{rev2}%
  \BibitemOpen
  \bibfield  {author} {\bibinfo {author} {\bibfnamefont {A.}~\bibnamefont {Polkovnikov}}, \bibinfo {author} {\bibfnamefont {K.}~\bibnamefont {Sengupta}}, \bibinfo {author} {\bibfnamefont {A.}~\bibnamefont {Silva}},\ and\ \bibinfo {author} {\bibfnamefont {M.}~\bibnamefont {Vengalattore}},\ }\bibfield  {title} {\bibinfo {title} {Colloquium: Nonequilibrium dynamics of closed interacting quantum systems},\ }\href {https://doi.org/10.1103/RevModPhys.83.863} {\bibfield  {journal} {\bibinfo  {journal} {Rev. Mod. Phys.}\ }\textbf {\bibinfo {volume} {83}},\ \bibinfo {pages} {863} (\bibinfo {year} {2011})}\BibitemShut {NoStop}%
\bibitem [{\citenamefont {Bukov}\ \emph {et~al.}(2015)\citenamefont {Bukov}, \citenamefont {D'Alessio},\ and\ \citenamefont {Polkovnikov}}]{rev3}%
  \BibitemOpen
  \bibfield  {author} {\bibinfo {author} {\bibfnamefont {M.}~\bibnamefont {Bukov}}, \bibinfo {author} {\bibfnamefont {L.}~\bibnamefont {D'Alessio}},\ and\ \bibinfo {author} {\bibfnamefont {A.}~\bibnamefont {Polkovnikov}},\ }\bibfield  {title} {\bibinfo {title} {Universal high-frequency behavior of periodically driven systems: from dynamical stabilization to floquet engineering},\ }\href {https://doi.org/10.1080/00018732.2015.1055918} {\bibfield  {journal} {\bibinfo  {journal} {Adv. Phys.}\ }\textbf {\bibinfo {volume} {64}},\ \bibinfo {pages} {139} (\bibinfo {year} {2015})}\BibitemShut {NoStop}%
\bibitem [{\citenamefont {Sen}\ \emph {et~al.}(2021)\citenamefont {Sen}, \citenamefont {Sen},\ and\ \citenamefont {Sengupta}}]{rev8}%
  \BibitemOpen
  \bibfield  {author} {\bibinfo {author} {\bibfnamefont {A.}~\bibnamefont {Sen}}, \bibinfo {author} {\bibfnamefont {D.}~\bibnamefont {Sen}},\ and\ \bibinfo {author} {\bibfnamefont {K.}~\bibnamefont {Sengupta}},\ }\bibfield  {title} {\bibinfo {title} {Analytic approaches to periodically driven closed quantum systems: methods and applications},\ }\href {https://doi.org/10.1088/1361-648X/ac1b61} {\bibfield  {journal} {\bibinfo  {journal} {J. Phys.: Condens. Matter}\ }\textbf {\bibinfo {volume} {33}},\ \bibinfo {pages} {443003} (\bibinfo {year} {2021})}\BibitemShut {NoStop}%
\bibitem [{\citenamefont {Bloch}\ \emph {et~al.}(2008)\citenamefont {Bloch}, \citenamefont {Dalibard},\ and\ \citenamefont {Zwerger}}]{rev9}%
  \BibitemOpen
  \bibfield  {author} {\bibinfo {author} {\bibfnamefont {I.}~\bibnamefont {Bloch}}, \bibinfo {author} {\bibfnamefont {J.}~\bibnamefont {Dalibard}},\ and\ \bibinfo {author} {\bibfnamefont {W.}~\bibnamefont {Zwerger}},\ }\bibfield  {title} {\bibinfo {title} {Many-body physics with ultracold gases},\ }\href {https://doi.org/10.1103/RevModPhys.80.885} {\bibfield  {journal} {\bibinfo  {journal} {Rev. Mod. Phys.}\ }\textbf {\bibinfo {volume} {80}},\ \bibinfo {pages} {885} (\bibinfo {year} {2008})}\BibitemShut {NoStop}%
\bibitem [{\citenamefont {Joshi}\ \emph {et~al.}(2023)\citenamefont {Joshi}, \citenamefont {Kokail}, \citenamefont {van Bijnen}, \citenamefont {Kranzl}, \citenamefont {Zache}, \citenamefont {Blatt}, \citenamefont {Roos},\ and\ \citenamefont {Zoller}}]{joshi23}%
  \BibitemOpen
  \bibfield  {author} {\bibinfo {author} {\bibfnamefont {M.~K.}\ \bibnamefont {Joshi}}, \bibinfo {author} {\bibfnamefont {C.}~\bibnamefont {Kokail}}, \bibinfo {author} {\bibfnamefont {R.}~\bibnamefont {van Bijnen}}, \bibinfo {author} {\bibfnamefont {F.}~\bibnamefont {Kranzl}}, \bibinfo {author} {\bibfnamefont {T.~V.}\ \bibnamefont {Zache}}, \bibinfo {author} {\bibfnamefont {R.}~\bibnamefont {Blatt}}, \bibinfo {author} {\bibfnamefont {C.~F.}\ \bibnamefont {Roos}},\ and\ \bibinfo {author} {\bibfnamefont {P.}~\bibnamefont {Zoller}},\ }\bibfield  {title} {\bibinfo {title} {Exploring large-scale entanglement in quantum simulation},\ }\href {https://doi.org/10.1038/s41586-023-06768-0} {\bibfield  {journal} {\bibinfo  {journal} {Nature}\ }\textbf {\bibinfo {volume} {624}},\ \bibinfo {pages} {539} (\bibinfo {year} {2023})}\BibitemShut {NoStop}%
\bibitem [{\citenamefont {Sun}\ \emph {et~al.}(2026)\citenamefont {Sun}, \citenamefont {Le}, \citenamefont {Naus} \emph {et~al.}}]{sun2026cftexp}%
  \BibitemOpen
  \bibfield  {author} {\bibinfo {author} {\bibfnamefont {X.}~\bibnamefont {Sun}}, \bibinfo {author} {\bibfnamefont {Y.}~\bibnamefont {Le}}, \bibinfo {author} {\bibfnamefont {S.}~\bibnamefont {Naus}}, \emph {et~al.},\ }\href@noop {} {\bibinfo {title} {Experimental observation of conformal field theory spectra}} (\bibinfo {year} {2026}),\ \Eprint {https://arxiv.org/abs/2601.16275} {arXiv:2601.16275} \BibitemShut {NoStop}%
\bibitem [{\citenamefont {Mo}\ \emph {et~al.}(2026)\citenamefont {Mo}, \citenamefont {Lapierre},\ and\ \citenamefont {Miao}}]{mo2026observingconformalfloquetdynamics}%
  \BibitemOpen
  \bibfield  {author} {\bibinfo {author} {\bibfnamefont {L.-H.}\ \bibnamefont {Mo}}, \bibinfo {author} {\bibfnamefont {B.}~\bibnamefont {Lapierre}},\ and\ \bibinfo {author} {\bibfnamefont {Q.}~\bibnamefont {Miao}},\ }\href@noop {} {\bibinfo {title} {{Observing conformal Floquet dynamics on a digital quantum processor}}} (\bibinfo {year} {2026}),\ \Eprint {https://arxiv.org/abs/2605.27530} {arXiv:2605.27530} \BibitemShut {NoStop}%
\bibitem [{\citenamefont {Wen}\ and\ \citenamefont {Wu}(2018{\natexlab{a}})}]{2dpapers1}%
  \BibitemOpen
  \bibfield  {author} {\bibinfo {author} {\bibfnamefont {X.}~\bibnamefont {Wen}}\ and\ \bibinfo {author} {\bibfnamefont {J.-Q.}\ \bibnamefont {Wu}},\ }\href@noop {} {\bibinfo {title} {Floquet conformal field theory}} (\bibinfo {year} {2018}{\natexlab{a}}),\ \Eprint {https://arxiv.org/abs/1805.00031} {arXiv:1805.00031} \BibitemShut {NoStop}%
\bibitem [{\citenamefont {Wen}\ and\ \citenamefont {Wu}(2018{\natexlab{b}})}]{dynph1}%
  \BibitemOpen
  \bibfield  {author} {\bibinfo {author} {\bibfnamefont {X.}~\bibnamefont {Wen}}\ and\ \bibinfo {author} {\bibfnamefont {J.-Q.}\ \bibnamefont {Wu}},\ }\bibfield  {title} {\bibinfo {title} {Quantum dynamics in sine-square deformed conformal field theory: Quench from uniform to nonuniform conformal field theory},\ }\href {https://doi.org/10.1103/PhysRevB.97.184309} {\bibfield  {journal} {\bibinfo  {journal} {Phys. Rev. B}\ }\textbf {\bibinfo {volume} {97}},\ \bibinfo {pages} {184309} (\bibinfo {year} {2018}{\natexlab{b}})}\BibitemShut {NoStop}%
\bibitem [{\citenamefont {Wen}\ \emph {et~al.}(2021)\citenamefont {Wen}, \citenamefont {Fan}, \citenamefont {Vishwanath},\ and\ \citenamefont {Gu}}]{dynph2}%
  \BibitemOpen
  \bibfield  {author} {\bibinfo {author} {\bibfnamefont {X.}~\bibnamefont {Wen}}, \bibinfo {author} {\bibfnamefont {R.}~\bibnamefont {Fan}}, \bibinfo {author} {\bibfnamefont {A.}~\bibnamefont {Vishwanath}},\ and\ \bibinfo {author} {\bibfnamefont {Y.}~\bibnamefont {Gu}},\ }\bibfield  {title} {\bibinfo {title} {Periodically, quasiperiodically, and randomly driven conformal field theories},\ }\href {https://doi.org/10.1103/PhysRevResearch.3.023044} {\bibfield  {journal} {\bibinfo  {journal} {Phys. Rev. Res.}\ }\textbf {\bibinfo {volume} {3}},\ \bibinfo {pages} {023044} (\bibinfo {year} {2021})}\BibitemShut {NoStop}%
\bibitem [{\citenamefont {Lapierre}\ \emph {et~al.}(2020{\natexlab{a}})\citenamefont {Lapierre}, \citenamefont {Choo}, \citenamefont {Tauber}, \citenamefont {Tiwari}, \citenamefont {Neupert},\ and\ \citenamefont {Chitra}}]{2dpapers2}%
  \BibitemOpen
  \bibfield  {author} {\bibinfo {author} {\bibfnamefont {B.}~\bibnamefont {Lapierre}}, \bibinfo {author} {\bibfnamefont {K.}~\bibnamefont {Choo}}, \bibinfo {author} {\bibfnamefont {C.}~\bibnamefont {Tauber}}, \bibinfo {author} {\bibfnamefont {A.}~\bibnamefont {Tiwari}}, \bibinfo {author} {\bibfnamefont {T.}~\bibnamefont {Neupert}},\ and\ \bibinfo {author} {\bibfnamefont {R.}~\bibnamefont {Chitra}},\ }\bibfield  {title} {\bibinfo {title} {Emergent black hole dynamics in critical floquet systems},\ }\href {https://doi.org/10.1103/PhysRevResearch.2.023085} {\bibfield  {journal} {\bibinfo  {journal} {Phys. Rev. Res.}\ }\textbf {\bibinfo {volume} {2}},\ \bibinfo {pages} {023085} (\bibinfo {year} {2020}{\natexlab{a}})}\BibitemShut {NoStop}%
\bibitem [{\citenamefont {Fan}\ \emph {et~al.}(2021)\citenamefont {Fan}, \citenamefont {Gu}, \citenamefont {Vishwanath},\ and\ \citenamefont {Wen}}]{2dpapers3}%
  \BibitemOpen
  \bibfield  {author} {\bibinfo {author} {\bibfnamefont {R.}~\bibnamefont {Fan}}, \bibinfo {author} {\bibfnamefont {Y.}~\bibnamefont {Gu}}, \bibinfo {author} {\bibfnamefont {A.}~\bibnamefont {Vishwanath}},\ and\ \bibinfo {author} {\bibfnamefont {X.}~\bibnamefont {Wen}},\ }\bibfield  {title} {\bibinfo {title} {Floquet conformal field theories with generally deformed hamiltonians},\ }\href {https://doi.org/10.21468/SciPostPhys.10.2.049} {\bibfield  {journal} {\bibinfo  {journal} {SciPost Phys.}\ }\textbf {\bibinfo {volume} {10}},\ \bibinfo {pages} {049} (\bibinfo {year} {2021})}\BibitemShut {NoStop}%
\bibitem [{\citenamefont {Das}\ \emph {et~al.}(2021)\citenamefont {Das}, \citenamefont {Ghosh},\ and\ \citenamefont {Sengupta}}]{2dpapers4}%
  \BibitemOpen
  \bibfield  {author} {\bibinfo {author} {\bibfnamefont {D.}~\bibnamefont {Das}}, \bibinfo {author} {\bibfnamefont {R.}~\bibnamefont {Ghosh}},\ and\ \bibinfo {author} {\bibfnamefont {K.}~\bibnamefont {Sengupta}},\ }\bibfield  {title} {\bibinfo {title} {{Conformal Floquet dynamics with a continuous drive protocol}},\ }\href {https://doi.org/10.1007/JHEP05(2021)172} {\bibfield  {journal} {\bibinfo  {journal} {{J. High Energy Phys.}}\ }\textbf {\bibinfo {volume} {05}},\ \bibinfo {pages} {172} (\bibinfo {year} {2021})}\BibitemShut {NoStop}%
\bibitem [{\citenamefont {Lapierre}\ \emph {et~al.}(2020{\natexlab{b}})\citenamefont {Lapierre}, \citenamefont {Choo}, \citenamefont {Tiwari}, \citenamefont {Tauber}, \citenamefont {Neupert},\ and\ \citenamefont {Chitra}}]{2dpapers5}%
  \BibitemOpen
  \bibfield  {author} {\bibinfo {author} {\bibfnamefont {B.}~\bibnamefont {Lapierre}}, \bibinfo {author} {\bibfnamefont {K.}~\bibnamefont {Choo}}, \bibinfo {author} {\bibfnamefont {A.}~\bibnamefont {Tiwari}}, \bibinfo {author} {\bibfnamefont {C.}~\bibnamefont {Tauber}}, \bibinfo {author} {\bibfnamefont {T.}~\bibnamefont {Neupert}},\ and\ \bibinfo {author} {\bibfnamefont {R.}~\bibnamefont {Chitra}},\ }\bibfield  {title} {\bibinfo {title} {{Fine structure of heating in a quasiperiodically driven critical quantum system}},\ }\href {https://doi.org/10.1103/PhysRevResearch.2.033461} {\bibfield  {journal} {\bibinfo  {journal} {Phys. Rev. Res.}\ }\textbf {\bibinfo {volume} {2}},\ \bibinfo {pages} {033461} (\bibinfo {year} {2020}{\natexlab{b}})}\BibitemShut {NoStop}%
\bibitem [{\citenamefont {Han}\ and\ \citenamefont {Wen}(2020)}]{2dpapers6}%
  \BibitemOpen
  \bibfield  {author} {\bibinfo {author} {\bibfnamefont {B.}~\bibnamefont {Han}}\ and\ \bibinfo {author} {\bibfnamefont {X.}~\bibnamefont {Wen}},\ }\bibfield  {title} {\bibinfo {title} {{Classification of $SL_2$ deformed Floquet conformal field theories}},\ }\href {https://doi.org/10.1103/PhysRevB.102.205125} {\bibfield  {journal} {\bibinfo  {journal} {Phys. Rev. B}\ }\textbf {\bibinfo {volume} {102}},\ \bibinfo {pages} {205125} (\bibinfo {year} {2020})}\BibitemShut {NoStop}%
\bibitem [{\citenamefont {Lapierre}\ \emph {et~al.}(2025{\natexlab{a}})\citenamefont {Lapierre}, \citenamefont {Numasawa}, \citenamefont {Neupert},\ and\ \citenamefont {Ryu}}]{2dpapers7}%
  \BibitemOpen
  \bibfield  {author} {\bibinfo {author} {\bibfnamefont {B.}~\bibnamefont {Lapierre}}, \bibinfo {author} {\bibfnamefont {T.}~\bibnamefont {Numasawa}}, \bibinfo {author} {\bibfnamefont {T.}~\bibnamefont {Neupert}},\ and\ \bibinfo {author} {\bibfnamefont {S.}~\bibnamefont {Ryu}},\ }\bibfield  {title} {\bibinfo {title} {{Floquet engineered inhomogeneous quantum chaos in critical systems}},\ }\href {https://doi.org/10.1103/cn3z-vfgr} {\bibfield  {journal} {\bibinfo  {journal} {Phys. Rev. B}\ }\textbf {\bibinfo {volume} {112}},\ \bibinfo {pages} {104317} (\bibinfo {year} {2025}{\natexlab{a}})}\BibitemShut {NoStop}%
\bibitem [{\citenamefont {Fang}\ \emph {et~al.}(2025)\citenamefont {Fang}, \citenamefont {Zhou},\ and\ \citenamefont {Wen}}]{2dpapers9}%
  \BibitemOpen
  \bibfield  {author} {\bibinfo {author} {\bibfnamefont {J.}~\bibnamefont {Fang}}, \bibinfo {author} {\bibfnamefont {Q.}~\bibnamefont {Zhou}},\ and\ \bibinfo {author} {\bibfnamefont {X.}~\bibnamefont {Wen}},\ }\bibfield  {title} {\bibinfo {title} {{Phase transitions in quasiperiodically driven quantum critical systems: Analytical results}},\ }\href {https://doi.org/https://doi.org/10.1103/PhysRevB.111.094304} {\bibfield  {journal} {\bibinfo  {journal} {Phys. Rev. B}\ }\textbf {\bibinfo {volume} {111}},\ \bibinfo {pages} {094304} (\bibinfo {year} {2025})}\BibitemShut {NoStop}%
\bibitem [{\citenamefont {Lapierre}\ \emph {et~al.}(2025{\natexlab{b}})\citenamefont {Lapierre}, \citenamefont {Pelliconi}, \citenamefont {Ryu},\ and\ \citenamefont {Sonner}}]{2dpapers10}%
  \BibitemOpen
  \bibfield  {author} {\bibinfo {author} {\bibfnamefont {B.}~\bibnamefont {Lapierre}}, \bibinfo {author} {\bibfnamefont {P.}~\bibnamefont {Pelliconi}}, \bibinfo {author} {\bibfnamefont {S.}~\bibnamefont {Ryu}},\ and\ \bibinfo {author} {\bibfnamefont {J.}~\bibnamefont {Sonner}},\ }\bibfield  {title} {\bibinfo {title} {{Driven nonunitary dynamics of quantum critical systems}},\ }\href {https://doi.org/10.1103/lwrz-jxrr} {\bibfield  {journal} {\bibinfo  {journal} {Phys. Rev. B}\ }\textbf {\bibinfo {volume} {112}},\ \bibinfo {pages} {104322} (\bibinfo {year} {2025}{\natexlab{b}})}\BibitemShut {NoStop}%
\bibitem [{\citenamefont {Lapierre}\ \emph {et~al.}(2025{\natexlab{c}})\citenamefont {Lapierre}, \citenamefont {Mo},\ and\ \citenamefont {Ryu}}]{2dpapers11}%
  \BibitemOpen
  \bibfield  {author} {\bibinfo {author} {\bibfnamefont {B.}~\bibnamefont {Lapierre}}, \bibinfo {author} {\bibfnamefont {L.-H.}\ \bibnamefont {Mo}},\ and\ \bibinfo {author} {\bibfnamefont {S.}~\bibnamefont {Ryu}},\ }\href@noop {} {\bibinfo {title} {{Entanglement transitions in structured and random nonunitary Gaussian circuits}}} (\bibinfo {year} {2025}{\natexlab{c}}),\ \Eprint {https://arxiv.org/abs/2507.03768} {arXiv:2507.03768} \BibitemShut {NoStop}%
\bibitem [{\citenamefont {Dey}\ \emph {et~al.}(2026)\citenamefont {Dey}, \citenamefont {Dutta},\ and\ \citenamefont {Ezhuthachan}}]{2dpapers12}%
  \BibitemOpen
  \bibfield  {author} {\bibinfo {author} {\bibfnamefont {P.}~\bibnamefont {Dey}}, \bibinfo {author} {\bibfnamefont {S.}~\bibnamefont {Dutta}},\ and\ \bibinfo {author} {\bibfnamefont {B.}~\bibnamefont {Ezhuthachan}},\ }\href@noop {} {\bibinfo {title} {{Imprints of dynamical phases in semiclassical entanglement entropy in 2D CFT}}} (\bibinfo {year} {2026}),\ \Eprint {https://arxiv.org/abs/2606.17625} {arXiv:2606.17625 [hep-th]} \BibitemShut {NoStop}%
\bibitem [{\citenamefont {Das}\ \emph {et~al.}(2022)\citenamefont {Das}, \citenamefont {Ezhuthachan}, \citenamefont {Kundu}, \citenamefont {Porey}, \citenamefont {Roy},\ and\ \citenamefont {Sengupta}}]{Das:2022jrr}%
  \BibitemOpen
  \bibfield  {author} {\bibinfo {author} {\bibfnamefont {S.}~\bibnamefont {Das}}, \bibinfo {author} {\bibfnamefont {B.}~\bibnamefont {Ezhuthachan}}, \bibinfo {author} {\bibfnamefont {A.}~\bibnamefont {Kundu}}, \bibinfo {author} {\bibfnamefont {S.}~\bibnamefont {Porey}}, \bibinfo {author} {\bibfnamefont {B.}~\bibnamefont {Roy}},\ and\ \bibinfo {author} {\bibfnamefont {K.}~\bibnamefont {Sengupta}},\ }\bibfield  {title} {\bibinfo {title} {{Out-of-Time-Order correlators in driven conformal field theories}},\ }\href {https://doi.org/10.1007/JHEP08(2022)221} {\bibfield  {journal} {\bibinfo  {journal} {{J. High Energy Phys.}}\ }\textbf {\bibinfo {volume} {08}},\ \bibinfo {pages} {221} (\bibinfo {year} {2022})}\BibitemShut {NoStop}%
\bibitem [{\citenamefont {Caputa}\ and\ \citenamefont {MacCormack}(2021)}]{2dcurved1}%
  \BibitemOpen
  \bibfield  {author} {\bibinfo {author} {\bibfnamefont {P.}~\bibnamefont {Caputa}}\ and\ \bibinfo {author} {\bibfnamefont {I.}~\bibnamefont {MacCormack}},\ }\bibfield  {title} {\bibinfo {title} {{Geometry and Complexity of Path Integrals in Inhomogeneous CFTs}},\ }\href {https://doi.org/10.1007/JHEP01(2021)027} {\bibfield  {journal} {\bibinfo  {journal} {{J. High Energy Phys.}}\ }\textbf {\bibinfo {volume} {01}},\ \bibinfo {pages} {027} (\bibinfo {year} {2021})},\ \bibinfo {note} {[Erratum: JHEP 09, 109 (2022)]}\BibitemShut {NoStop}%
\bibitem [{\citenamefont {de~Boer}\ \emph {et~al.}(2023)\citenamefont {de~Boer}, \citenamefont {Godet}, \citenamefont {Kastikainen},\ and\ \citenamefont {Keski-Vakkuri}}]{2dcurved2}%
  \BibitemOpen
  \bibfield  {author} {\bibinfo {author} {\bibfnamefont {J.}~\bibnamefont {de~Boer}}, \bibinfo {author} {\bibfnamefont {V.}~\bibnamefont {Godet}}, \bibinfo {author} {\bibfnamefont {J.}~\bibnamefont {Kastikainen}},\ and\ \bibinfo {author} {\bibfnamefont {E.}~\bibnamefont {Keski-Vakkuri}},\ }\bibfield  {title} {\bibinfo {title} {{Quantum information geometry of driven CFTs}},\ }\href {https://doi.org/10.1007/JHEP09(2023)087} {\bibfield  {journal} {\bibinfo  {journal} {{J. High Energy Phys.}}\ }\textbf {\bibinfo {volume} {09}},\ \bibinfo {pages} {087} (\bibinfo {year} {2023})}\BibitemShut {NoStop}%
\bibitem [{\citenamefont {Erdmenger}\ \emph {et~al.}(2026)\citenamefont {Erdmenger}, \citenamefont {Kastikainen},\ and\ \citenamefont {Schuhmann}}]{2dcurved3}%
  \BibitemOpen
  \bibfield  {author} {\bibinfo {author} {\bibfnamefont {J.}~\bibnamefont {Erdmenger}}, \bibinfo {author} {\bibfnamefont {J.}~\bibnamefont {Kastikainen}},\ and\ \bibinfo {author} {\bibfnamefont {T.}~\bibnamefont {Schuhmann}},\ }\bibfield  {title} {\bibinfo {title} {{Driven inhomogeneous CFT as a theory in curved space-time}},\ }\href {https://doi.org/10.1007/JHEP02(2026)255} {\bibfield  {journal} {\bibinfo  {journal} {{J. High Energy Phys.}}\ }\textbf {\bibinfo {volume} {02}},\ \bibinfo {pages} {255} (\bibinfo {year} {2026})}\BibitemShut {NoStop}%
\bibitem [{\citenamefont {MacCormack}\ \emph {et~al.}(2019)\citenamefont {MacCormack}, \citenamefont {Liu}, \citenamefont {Nozaki},\ and\ \citenamefont {Ryu}}]{holo1}%
  \BibitemOpen
  \bibfield  {author} {\bibinfo {author} {\bibfnamefont {I.}~\bibnamefont {MacCormack}}, \bibinfo {author} {\bibfnamefont {A.}~\bibnamefont {Liu}}, \bibinfo {author} {\bibfnamefont {M.}~\bibnamefont {Nozaki}},\ and\ \bibinfo {author} {\bibfnamefont {S.}~\bibnamefont {Ryu}},\ }\bibfield  {title} {\bibinfo {title} {{Holographic Duals of Inhomogeneous Systems: The Rainbow Chain and the Sine-Square Deformation Model}},\ }\href {https://doi.org/10.1088/1751-8121/ab3944} {\bibfield  {journal} {\bibinfo  {journal} {J. Phys. A}\ }\textbf {\bibinfo {volume} {52}},\ \bibinfo {pages} {505401} (\bibinfo {year} {2019})}\BibitemShut {NoStop}%
\bibitem [{\citenamefont {Das}\ \emph {et~al.}(2023)\citenamefont {Das}, \citenamefont {Ezhuthachan}, \citenamefont {Kundu}, \citenamefont {Porey}, \citenamefont {Roy},\ and\ \citenamefont {Sengupta}}]{holo2}%
  \BibitemOpen
  \bibfield  {author} {\bibinfo {author} {\bibfnamefont {S.}~\bibnamefont {Das}}, \bibinfo {author} {\bibfnamefont {B.}~\bibnamefont {Ezhuthachan}}, \bibinfo {author} {\bibfnamefont {A.}~\bibnamefont {Kundu}}, \bibinfo {author} {\bibfnamefont {S.}~\bibnamefont {Porey}}, \bibinfo {author} {\bibfnamefont {B.}~\bibnamefont {Roy}},\ and\ \bibinfo {author} {\bibfnamefont {K.}~\bibnamefont {Sengupta}},\ }\bibfield  {title} {\bibinfo {title} {{Brane detectors of a dynamical phase transition in a driven CFT}},\ }\href {https://doi.org/10.21468/SciPostPhys.15.5.202} {\bibfield  {journal} {\bibinfo  {journal} {SciPost Phys.}\ }\textbf {\bibinfo {volume} {15}},\ \bibinfo {pages} {202} (\bibinfo {year} {2023})}\BibitemShut {NoStop}%
\bibitem [{\citenamefont {Kudler-Flam}\ \emph {et~al.}(2024)\citenamefont {Kudler-Flam}, \citenamefont {Nozaki}, \citenamefont {Numasawa}, \citenamefont {Ryu},\ and\ \citenamefont {Tan}}]{holo3}%
  \BibitemOpen
  \bibfield  {author} {\bibinfo {author} {\bibfnamefont {J.}~\bibnamefont {Kudler-Flam}}, \bibinfo {author} {\bibfnamefont {M.}~\bibnamefont {Nozaki}}, \bibinfo {author} {\bibfnamefont {T.}~\bibnamefont {Numasawa}}, \bibinfo {author} {\bibfnamefont {S.}~\bibnamefont {Ryu}},\ and\ \bibinfo {author} {\bibfnamefont {M.~T.}\ \bibnamefont {Tan}},\ }\bibfield  {title} {\bibinfo {title} {{Bridging two quantum quench problems {\textemdash} local joining quantum quench and M{\"o}bius quench {\textemdash} and their holographic dual descriptions}},\ }\href {https://doi.org/10.1007/JHEP08(2024)213} {\bibfield  {journal} {\bibinfo  {journal} {{J. High Energy Phys.}}\ }\textbf {\bibinfo {volume} {08}},\ \bibinfo {pages} {213} (\bibinfo {year} {2024})}\BibitemShut {NoStop}%
\bibitem [{\citenamefont {Jiang}\ and\ \citenamefont {Mezei}(2025)}]{holo4}%
  \BibitemOpen
  \bibfield  {author} {\bibinfo {author} {\bibfnamefont {H.}~\bibnamefont {Jiang}}\ and\ \bibinfo {author} {\bibfnamefont {M.}~\bibnamefont {Mezei}},\ }\bibfield  {title} {\bibinfo {title} {{New horizons for inhomogeneous quenches and Floquet CFT}},\ }\href {https://doi.org/10.1007/JHEP04(2025)025} {\bibfield  {journal} {\bibinfo  {journal} {{J. High Energy Phys.}}\ }\textbf {\bibinfo {volume} {04}},\ \bibinfo {pages} {025} (\bibinfo {year} {2025})}\BibitemShut {NoStop}%
\bibitem [{\citenamefont {Das}\ and\ \citenamefont {Kundu}(2025)}]{holo5}%
  \BibitemOpen
  \bibfield  {author} {\bibinfo {author} {\bibfnamefont {J.}~\bibnamefont {Das}}\ and\ \bibinfo {author} {\bibfnamefont {A.}~\bibnamefont {Kundu}},\ }\bibfield  {title} {\bibinfo {title} {{Flowery horizons {\&} bulk observers: sl$^{(q)}$(2,{\,}{\ensuremath{\mathbb{R}}}) drive in 2d holographic CFT}},\ }\href {https://doi.org/10.1007/JHEP05(2025)035} {\bibfield  {journal} {\bibinfo  {journal} {{J. High Energy Phys.}}\ }\textbf {\bibinfo {volume} {05}},\ \bibinfo {pages} {035} (\bibinfo {year} {2025})}\BibitemShut {NoStop}%
\bibitem [{\citenamefont {Das}\ \emph {et~al.}(2024)\citenamefont {Das}, \citenamefont {Das}, \citenamefont {Kundu},\ and\ \citenamefont {Sengupta}}]{hd1}%
  \BibitemOpen
  \bibfield  {author} {\bibinfo {author} {\bibfnamefont {D.}~\bibnamefont {Das}}, \bibinfo {author} {\bibfnamefont {S.~R.}\ \bibnamefont {Das}}, \bibinfo {author} {\bibfnamefont {A.}~\bibnamefont {Kundu}},\ and\ \bibinfo {author} {\bibfnamefont {K.}~\bibnamefont {Sengupta}},\ }\bibfield  {title} {\bibinfo {title} {{Exactly solvable Floquet dynamics for conformal field theories in dimensions greater than two}},\ }\href {https://doi.org/10.1007/JHEP09(2024)095} {\bibfield  {journal} {\bibinfo  {journal} {{J. High Energy Phys.}}\ }\textbf {\bibinfo {volume} {09}},\ \bibinfo {pages} {95} (\bibinfo {year} {2024})}\BibitemShut {NoStop}%
\bibitem [{\citenamefont {Das}\ \emph {et~al.}(2026)\citenamefont {Das}, \citenamefont {Das}, \citenamefont {Kundu},\ and\ \citenamefont {Sengupta}}]{hd2}%
  \BibitemOpen
  \bibfield  {author} {\bibinfo {author} {\bibfnamefont {D.}~\bibnamefont {Das}}, \bibinfo {author} {\bibfnamefont {S.~R.}\ \bibnamefont {Das}}, \bibinfo {author} {\bibfnamefont {A.}~\bibnamefont {Kundu}},\ and\ \bibinfo {author} {\bibfnamefont {K.}~\bibnamefont {Sengupta}},\ }\bibfield  {title} {\bibinfo {title} {{Dynamical phases of higher dimensional Floquet CFTs}},\ }\href {https://doi.org/10.21468/SciPostPhys.20.2.045} {\bibfield  {journal} {\bibinfo  {journal} {SciPost Phys.}\ }\textbf {\bibinfo {volume} {20}},\ \bibinfo {pages} {045} (\bibinfo {year} {2026})}\BibitemShut {NoStop}%
\bibitem [{\citenamefont {Ares}\ \emph {et~al.}(2023{\natexlab{a}})\citenamefont {Ares}, \citenamefont {Murciano},\ and\ \citenamefont {Calabrese}}]{ares23asymmetry}%
  \BibitemOpen
  \bibfield  {author} {\bibinfo {author} {\bibfnamefont {F.}~\bibnamefont {Ares}}, \bibinfo {author} {\bibfnamefont {S.}~\bibnamefont {Murciano}},\ and\ \bibinfo {author} {\bibfnamefont {P.}~\bibnamefont {Calabrese}},\ }\bibfield  {title} {\bibinfo {title} {Entanglement asymmetry as a probe of symmetry breaking},\ }\href {https://doi.org/https://doi.org/10.1038/s41467-023-37747-8} {\bibfield  {journal} {\bibinfo  {journal} {Nat. Comms.}\ }\textbf {\bibinfo {volume} {14}},\ \bibinfo {pages} {2036} (\bibinfo {year} {2023}{\natexlab{a}})}\BibitemShut {NoStop}%
\bibitem [{\citenamefont {Bartlett}\ \emph {et~al.}(2007)\citenamefont {Bartlett}, \citenamefont {Rudolph},\ and\ \citenamefont {Spekkens}}]{bartlett07}%
  \BibitemOpen
  \bibfield  {author} {\bibinfo {author} {\bibfnamefont {S.~D.}\ \bibnamefont {Bartlett}}, \bibinfo {author} {\bibfnamefont {T.}~\bibnamefont {Rudolph}},\ and\ \bibinfo {author} {\bibfnamefont {R.~W.}\ \bibnamefont {Spekkens}},\ }\bibfield  {title} {\bibinfo {title} {Reference frames, superselection rules, and quantum information},\ }\href {https://doi.org/10.1103/RevModPhys.79.555} {\bibfield  {journal} {\bibinfo  {journal} {Rev. Mod. Phys.}\ }\textbf {\bibinfo {volume} {79}},\ \bibinfo {pages} {555} (\bibinfo {year} {2007})}\BibitemShut {NoStop}%
\bibitem [{\citenamefont {Vaccaro}\ \emph {et~al.}(2008)\citenamefont {Vaccaro}, \citenamefont {Anselmi}, \citenamefont {Wiseman},\ and\ \citenamefont {Jacobs}}]{vaccaro08}%
  \BibitemOpen
  \bibfield  {author} {\bibinfo {author} {\bibfnamefont {J.~A.}\ \bibnamefont {Vaccaro}}, \bibinfo {author} {\bibfnamefont {F.}~\bibnamefont {Anselmi}}, \bibinfo {author} {\bibfnamefont {H.~M.}\ \bibnamefont {Wiseman}},\ and\ \bibinfo {author} {\bibfnamefont {K.}~\bibnamefont {Jacobs}},\ }\bibfield  {title} {\bibinfo {title} {Tradeoff between extractable mechanical work, accessible entanglement, and ability to act as a reference system, under arbitrary superselection rules},\ }\href {https://doi.org/10.1103/PhysRevA.77.032114} {\bibfield  {journal} {\bibinfo  {journal} {Phys. Rev. A}\ }\textbf {\bibinfo {volume} {77}},\ \bibinfo {pages} {032114} (\bibinfo {year} {2008})}\BibitemShut {NoStop}%
\bibitem [{\citenamefont {Gour}\ \emph {et~al.}(2009)\citenamefont {Gour}, \citenamefont {Marvian},\ and\ \citenamefont {Spekkens}}]{gour09}%
  \BibitemOpen
  \bibfield  {author} {\bibinfo {author} {\bibfnamefont {G.}~\bibnamefont {Gour}}, \bibinfo {author} {\bibfnamefont {I.}~\bibnamefont {Marvian}},\ and\ \bibinfo {author} {\bibfnamefont {R.~W.}\ \bibnamefont {Spekkens}},\ }\bibfield  {title} {\bibinfo {title} {{Measuring the quality of a quantum reference frame: The relative entropy of frameness}},\ }\href {https://doi.org/10.1103/PhysRevA.80.012307} {\bibfield  {journal} {\bibinfo  {journal} {Phys. Rev. A}\ }\textbf {\bibinfo {volume} {80}},\ \bibinfo {pages} {012307} (\bibinfo {year} {2009})}\BibitemShut {NoStop}%
\bibitem [{\citenamefont {Marvian}\ and\ \citenamefont {Spekkens}(2014)}]{marvian14}%
  \BibitemOpen
  \bibfield  {author} {\bibinfo {author} {\bibfnamefont {I.}~\bibnamefont {Marvian}}\ and\ \bibinfo {author} {\bibfnamefont {R.~W.}\ \bibnamefont {Spekkens}},\ }\bibfield  {title} {\bibinfo {title} {{Extending Noether’s theorem by quantifying the asymmetry of quantum states}},\ }\href {https://doi.org/10.1038/ncomms4821} {\bibfield  {journal} {\bibinfo  {journal} {Nat. Comms.}\ }\textbf {\bibinfo {volume} {5}},\ \bibinfo {pages} {3821} (\bibinfo {year} {2014})}\BibitemShut {NoStop}%
\bibitem [{\citenamefont {Joshi}\ \emph {et~al.}(2024)\citenamefont {Joshi}, \citenamefont {Franke}, \citenamefont {Rath} \emph {et~al.}}]{joshi24}%
  \BibitemOpen
  \bibfield  {author} {\bibinfo {author} {\bibfnamefont {L.~K.}\ \bibnamefont {Joshi}}, \bibinfo {author} {\bibfnamefont {J.}~\bibnamefont {Franke}}, \bibinfo {author} {\bibfnamefont {A.}~\bibnamefont {Rath}}, \emph {et~al.},\ }\bibfield  {title} {\bibinfo {title} {{Observing the Quantum Mpemba Effect in Quantum Simulations}},\ }\href {https://doi.org/10.1103/PhysRevLett.133.010402} {\bibfield  {journal} {\bibinfo  {journal} {Phys. Rev. Lett.}\ }\textbf {\bibinfo {volume} {133}},\ \bibinfo {pages} {010402} (\bibinfo {year} {2024})}\BibitemShut {NoStop}%
\bibitem [{\citenamefont {Xu}\ \emph {et~al.}(2026)\citenamefont {Xu}, \citenamefont {Fang}, \citenamefont {Chen} \emph {et~al.}}]{xu2026EAexp}%
  \BibitemOpen
  \bibfield  {author} {\bibinfo {author} {\bibfnamefont {Y.}~\bibnamefont {Xu}}, \bibinfo {author} {\bibfnamefont {C.-P.}\ \bibnamefont {Fang}}, \bibinfo {author} {\bibfnamefont {B.-J.}\ \bibnamefont {Chen}}, \emph {et~al.},\ }\bibfield  {title} {\bibinfo {title} {{Observation and Modulation of the Quantum Mpemba Effect on a Superconducting Quantum Processor}},\ }\href {https://doi.org/10.1103/951q-j8kq} {\bibfield  {journal} {\bibinfo  {journal} {Phys. Rev. Lett.}\ }\textbf {\bibinfo {volume} {137}} (\bibinfo {year} {2026})}\BibitemShut {NoStop}%
\bibitem [{\citenamefont {Yang}\ \emph {et~al.}(2026)\citenamefont {Yang}, \citenamefont {Joshi}, \citenamefont {Ares}, \citenamefont {Han}, \citenamefont {Zhang},\ and\ \citenamefont {Calabrese}}]{Yang26Pageexp}%
  \BibitemOpen
  \bibfield  {author} {\bibinfo {author} {\bibfnamefont {J.-N.}\ \bibnamefont {Yang}}, \bibinfo {author} {\bibfnamefont {L.~K.}\ \bibnamefont {Joshi}}, \bibinfo {author} {\bibfnamefont {F.}~\bibnamefont {Ares}}, \bibinfo {author} {\bibfnamefont {Y.}~\bibnamefont {Han}}, \bibinfo {author} {\bibfnamefont {P.}~\bibnamefont {Zhang}},\ and\ \bibinfo {author} {\bibfnamefont {P.}~\bibnamefont {Calabrese}},\ }\href@noop {} {\bibinfo {title} {{Probing Entanglement and Symmetries in Random States Using a Superconducting Quantum Processor}}} (\bibinfo {year} {2026}),\ \Eprint {https://arxiv.org/abs/2601.22224} {arXiv:2601.22224} \BibitemShut {NoStop}%
\bibitem [{\citenamefont {Banerjee}\ \emph {et~al.}(2025)\citenamefont {Banerjee}, \citenamefont {Das},\ and\ \citenamefont {Sengupta}}]{Banerjee:2024zqb}%
  \BibitemOpen
  \bibfield  {author} {\bibinfo {author} {\bibfnamefont {T.}~\bibnamefont {Banerjee}}, \bibinfo {author} {\bibfnamefont {S.}~\bibnamefont {Das}},\ and\ \bibinfo {author} {\bibfnamefont {K.}~\bibnamefont {Sengupta}},\ }\bibfield  {title} {\bibinfo {title} {{Entanglement asymmetry in periodically driven quantum systems}},\ }\href {https://doi.org/10.21468/SciPostPhys.19.2.051} {\bibfield  {journal} {\bibinfo  {journal} {SciPost Phys.}\ }\textbf {\bibinfo {volume} {19}},\ \bibinfo {pages} {051} (\bibinfo {year} {2025})}\BibitemShut {NoStop}%
\bibitem [{\citenamefont {Benini}\ \emph {et~al.}(2025{\natexlab{a}})\citenamefont {Benini}, \citenamefont {Godet},\ and\ \citenamefont {Singh}}]{Benini:2024xjv}%
  \BibitemOpen
  \bibfield  {author} {\bibinfo {author} {\bibfnamefont {F.}~\bibnamefont {Benini}}, \bibinfo {author} {\bibfnamefont {V.}~\bibnamefont {Godet}},\ and\ \bibinfo {author} {\bibfnamefont {A.~H.}\ \bibnamefont {Singh}},\ }\bibfield  {title} {\bibinfo {title} {{Entanglement asymmetry in conformal field theory and holography}},\ }\href {https://doi.org/10.1093/ptep/ptaf080} {\bibfield  {journal} {\bibinfo  {journal} {Prog. Theor. Exp. Phys.}\ }\textbf {\bibinfo {volume} {6}},\ \bibinfo {pages} {063B05} (\bibinfo {year} {2025}{\natexlab{a}})}\BibitemShut {NoStop}%
\bibitem [{\citenamefont {Rylands}\ \emph {et~al.}(2024)\citenamefont {Rylands}, \citenamefont {Klobas}, \citenamefont {Ares}, \citenamefont {Calabrese}, \citenamefont {Murciano},\ and\ \citenamefont {Bertini}}]{rylands24}%
  \BibitemOpen
  \bibfield  {author} {\bibinfo {author} {\bibfnamefont {C.}~\bibnamefont {Rylands}}, \bibinfo {author} {\bibfnamefont {K.}~\bibnamefont {Klobas}}, \bibinfo {author} {\bibfnamefont {F.}~\bibnamefont {Ares}}, \bibinfo {author} {\bibfnamefont {P.}~\bibnamefont {Calabrese}}, \bibinfo {author} {\bibfnamefont {S.}~\bibnamefont {Murciano}},\ and\ \bibinfo {author} {\bibfnamefont {B.}~\bibnamefont {Bertini}},\ }\bibfield  {title} {\bibinfo {title} {{Microscopic Origin of the Quantum Mpemba Effect in Integrable Systems}},\ }\href {https://doi.org/10.1103/PhysRevLett.133.010401} {\bibfield  {journal} {\bibinfo  {journal} {Phys. Rev. Lett.}\ }\textbf {\bibinfo {volume} {133}},\ \bibinfo {pages} {010401} (\bibinfo {year} {2024})}\BibitemShut {NoStop}%
\bibitem [{\citenamefont {Murciano}\ \emph {et~al.}(2024)\citenamefont {Murciano}, \citenamefont {Ares}, \citenamefont {Klich},\ and\ \citenamefont {Calabrese}}]{Murciano2024}%
  \BibitemOpen
  \bibfield  {author} {\bibinfo {author} {\bibfnamefont {S.}~\bibnamefont {Murciano}}, \bibinfo {author} {\bibfnamefont {F.}~\bibnamefont {Ares}}, \bibinfo {author} {\bibfnamefont {I.}~\bibnamefont {Klich}},\ and\ \bibinfo {author} {\bibfnamefont {P.}~\bibnamefont {Calabrese}},\ }\bibfield  {title} {\bibinfo {title} {{Entanglement asymmetry and quantum Mpemba effect in the {XY} spin chain}},\ }\href {https://doi.org/10.1088/1742-5468/ad17b4} {\bibfield  {journal} {\bibinfo  {journal} {J. Stat. Mech}\ ,\ \bibinfo {pages} {013101}} (\bibinfo {year} {2024})}\BibitemShut {NoStop}%
\bibitem [{\citenamefont {Yamashika}\ \emph {et~al.}(2024)\citenamefont {Yamashika}, \citenamefont {Ares},\ and\ \citenamefont {Calabrese}}]{Yamashika_2024}%
  \BibitemOpen
  \bibfield  {author} {\bibinfo {author} {\bibfnamefont {S.}~\bibnamefont {Yamashika}}, \bibinfo {author} {\bibfnamefont {F.}~\bibnamefont {Ares}},\ and\ \bibinfo {author} {\bibfnamefont {P.}~\bibnamefont {Calabrese}},\ }\bibfield  {title} {\bibinfo {title} {{Entanglement asymmetry and quantum Mpemba effect in two-dimensional free-fermion systems}},\ }\href {https://doi.org/10.1103/physrevb.110.085126} {\bibfield  {journal} {\bibinfo  {journal} {Physical Review B}\ }\textbf {\bibinfo {volume} {110}},\ \bibinfo {pages} {085126} (\bibinfo {year} {2024})}\BibitemShut {NoStop}%
\bibitem [{\citenamefont {Ares}\ \emph {et~al.}(2025{\natexlab{a}})\citenamefont {Ares}, \citenamefont {Vitale},\ and\ \citenamefont {Murciano}}]{Ares_2025}%
  \BibitemOpen
  \bibfield  {author} {\bibinfo {author} {\bibfnamefont {F.}~\bibnamefont {Ares}}, \bibinfo {author} {\bibfnamefont {V.}~\bibnamefont {Vitale}},\ and\ \bibinfo {author} {\bibfnamefont {S.}~\bibnamefont {Murciano}},\ }\bibfield  {title} {\bibinfo {title} {{Quantum Mpemba effect in free-fermionic mixed states}},\ }\href {https://doi.org/10.1103/physrevb.111.104312} {\bibfield  {journal} {\bibinfo  {journal} {Phys. Rev. B}\ }\textbf {\bibinfo {volume} {111}},\ \bibinfo {pages} {104312} (\bibinfo {year} {2025}{\natexlab{a}})}\BibitemShut {NoStop}%
\bibitem [{\citenamefont {Liu}\ \emph {et~al.}(2024)\citenamefont {Liu}, \citenamefont {Zhang}, \citenamefont {Yin},\ and\ \citenamefont {Zhang}}]{liu24}%
  \BibitemOpen
  \bibfield  {author} {\bibinfo {author} {\bibfnamefont {S.}~\bibnamefont {Liu}}, \bibinfo {author} {\bibfnamefont {H.-K.}\ \bibnamefont {Zhang}}, \bibinfo {author} {\bibfnamefont {S.}~\bibnamefont {Yin}},\ and\ \bibinfo {author} {\bibfnamefont {S.-X.}\ \bibnamefont {Zhang}},\ }\bibfield  {title} {\bibinfo {title} {{Symmetry Restoration and Quantum Mpemba Effect in Symmetric Random Circuits}},\ }\href {https://doi.org/10.1103/PhysRevLett.133.140405} {\bibfield  {journal} {\bibinfo  {journal} {Phys. Rev. Lett.}\ }\textbf {\bibinfo {volume} {133}},\ \bibinfo {pages} {140405} (\bibinfo {year} {2024})}\BibitemShut {NoStop}%
\bibitem [{\citenamefont {Turkeshi}\ \emph {et~al.}(2025)\citenamefont {Turkeshi}, \citenamefont {Calabrese},\ and\ \citenamefont {De~Luca}}]{turkeshi25}%
  \BibitemOpen
  \bibfield  {author} {\bibinfo {author} {\bibfnamefont {X.}~\bibnamefont {Turkeshi}}, \bibinfo {author} {\bibfnamefont {P.}~\bibnamefont {Calabrese}},\ and\ \bibinfo {author} {\bibfnamefont {A.}~\bibnamefont {De~Luca}},\ }\bibfield  {title} {\bibinfo {title} {{Quantum Mpemba Effect in Random Circuits}},\ }\href {https://doi.org/10.1103/5d6p-8d1b} {\bibfield  {journal} {\bibinfo  {journal} {Phys. Rev. Lett.}\ }\textbf {\bibinfo {volume} {135}},\ \bibinfo {pages} {040403} (\bibinfo {year} {2025})}\BibitemShut {NoStop}%
\bibitem [{\citenamefont {Ares}\ \emph {et~al.}(2025{\natexlab{b}})\citenamefont {Ares}, \citenamefont {Murciano}, \citenamefont {Calabrese},\ and\ \citenamefont {Piroli}}]{ares25circ}%
  \BibitemOpen
  \bibfield  {author} {\bibinfo {author} {\bibfnamefont {F.}~\bibnamefont {Ares}}, \bibinfo {author} {\bibfnamefont {S.}~\bibnamefont {Murciano}}, \bibinfo {author} {\bibfnamefont {P.}~\bibnamefont {Calabrese}},\ and\ \bibinfo {author} {\bibfnamefont {L.}~\bibnamefont {Piroli}},\ }\bibfield  {title} {\bibinfo {title} {{Entanglement asymmetry dynamics in random quantum circuits}},\ }\href {https://doi.org/10.1103/m3np-p5xj} {\bibfield  {journal} {\bibinfo  {journal} {Phys. Rev. Research}\ }\textbf {\bibinfo {volume} {7}},\ \bibinfo {pages} {033135} (\bibinfo {year} {2025}{\natexlab{b}})}\BibitemShut {NoStop}%
\bibitem [{\citenamefont {Castro-Alvaredo}\ \emph {et~al.}(2026)\citenamefont {Castro-Alvaredo}, \citenamefont {Szász-Schagrin},\ and\ \citenamefont {Mazzoni}}]{castroalvaredo26}%
  \BibitemOpen
  \bibfield  {author} {\bibinfo {author} {\bibfnamefont {O.~A.}\ \bibnamefont {Castro-Alvaredo}}, \bibinfo {author} {\bibfnamefont {D.}~\bibnamefont {Szász-Schagrin}},\ and\ \bibinfo {author} {\bibfnamefont {M.}~\bibnamefont {Mazzoni}},\ }\href@noop {} {\bibinfo {title} {{Entanglement Asymmetry in Random Quantum Automata}}} (\bibinfo {year} {2026}),\ \Eprint {https://arxiv.org/abs/2607.07556} {arXiv:2607.07556} \BibitemShut {NoStop}%
\bibitem [{\citenamefont {Yamashika}\ and\ \citenamefont {Ares}(2026)}]{yamashika26}%
  \BibitemOpen
  \bibfield  {author} {\bibinfo {author} {\bibfnamefont {S.}~\bibnamefont {Yamashika}}\ and\ \bibinfo {author} {\bibfnamefont {F.}~\bibnamefont {Ares}},\ }\bibfield  {title} {\bibinfo {title} {{Quantum Mpemba Effect in Long-Range Spin Systems}},\ }\href {https://doi.org/10.1103/52y5-8kl2} {\bibfield  {journal} {\bibinfo  {journal} {Phys. Rev. Lett.}\ }\textbf {\bibinfo {volume} {136}},\ \bibinfo {pages} {090402} (\bibinfo {year} {2026})}\BibitemShut {NoStop}%
\bibitem [{\citenamefont {Russotto}\ \emph {et~al.}(2026)\citenamefont {Russotto}, \citenamefont {Ares}, \citenamefont {Calabrese},\ and\ \citenamefont {Alba}}]{russotto26}%
  \BibitemOpen
  \bibfield  {author} {\bibinfo {author} {\bibfnamefont {A.}~\bibnamefont {Russotto}}, \bibinfo {author} {\bibfnamefont {F.}~\bibnamefont {Ares}}, \bibinfo {author} {\bibfnamefont {P.}~\bibnamefont {Calabrese}},\ and\ \bibinfo {author} {\bibfnamefont {V.}~\bibnamefont {Alba}},\ }\bibfield  {title} {\bibinfo {title} {{Dynamics of entanglement fluctuations and quantum Mpemba effect in the $\nu=1$ QSSEP model}},\ }\href {https://iopscience.iop.org/article/10.1088/1742-5468/ae4bb9/meta} {\bibfield  {journal} {\bibinfo  {journal} {J. Stat. Mech.}\ ,\ \bibinfo {pages} {033103}} (\bibinfo {year} {2026})}\BibitemShut {NoStop}%
\bibitem [{\citenamefont {Summer}\ \emph {et~al.}(2026)\citenamefont {Summer}, \citenamefont {Moroder}, \citenamefont {Bettmann}, \citenamefont {Turkeshi}, \citenamefont {Marvian},\ and\ \citenamefont {Goold}}]{summer26}%
  \BibitemOpen
  \bibfield  {author} {\bibinfo {author} {\bibfnamefont {A.}~\bibnamefont {Summer}}, \bibinfo {author} {\bibfnamefont {M.}~\bibnamefont {Moroder}}, \bibinfo {author} {\bibfnamefont {L.~P.}\ \bibnamefont {Bettmann}}, \bibinfo {author} {\bibfnamefont {X.}~\bibnamefont {Turkeshi}}, \bibinfo {author} {\bibfnamefont {I.}~\bibnamefont {Marvian}},\ and\ \bibinfo {author} {\bibfnamefont {J.}~\bibnamefont {Goold}},\ }\bibfield  {title} {\bibinfo {title} {{Resource-Theoretical Unification of Mpemba Effects: Classical and Quantum}},\ }\href {https://doi.org/10.1103/rbt4-psfd} {\bibfield  {journal} {\bibinfo  {journal} {Phys. Rev. X}\ }\textbf {\bibinfo {volume} {16}},\ \bibinfo {pages} {011065} (\bibinfo {year} {2026})}\BibitemShut {NoStop}%
\bibitem [{\citenamefont {Müller}\ \emph {et~al.}(2026)\citenamefont {Müller}, \citenamefont {Pappalardi},\ and\ \citenamefont {Fazio}}]{muller26}%
  \BibitemOpen
  \bibfield  {author} {\bibinfo {author} {\bibfnamefont {T.~M.}\ \bibnamefont {Müller}}, \bibinfo {author} {\bibfnamefont {S.}~\bibnamefont {Pappalardi}},\ and\ \bibinfo {author} {\bibfnamefont {R.}~\bibnamefont {Fazio}},\ }\bibfield  {title} {\bibinfo {title} {{Quantum Mpemba effect in chaotic systems with conservation laws}},\ }\href {https://doi.org/10.1103/8nhn-1rs1} {\bibfield  {journal} {\bibinfo  {journal} {Phys. Rev. B}\ }\textbf {\bibinfo {volume} {114}},\ \bibinfo {pages} {L080301} (\bibinfo {year} {2026})}\BibitemShut {NoStop}%
\bibitem [{\citenamefont {Aditya}\ \emph {et~al.}(2026)\citenamefont {Aditya}, \citenamefont {Murciano},\ and\ \citenamefont {Turkeshi}}]{aditya26}%
  \BibitemOpen
  \bibfield  {author} {\bibinfo {author} {\bibfnamefont {S.}~\bibnamefont {Aditya}}, \bibinfo {author} {\bibfnamefont {S.}~\bibnamefont {Murciano}},\ and\ \bibinfo {author} {\bibfnamefont {X.}~\bibnamefont {Turkeshi}},\ }\href@noop {} {\bibinfo {title} {{Higher-order Symmetric Quantum Mpemba Effect in Fragmented Systems}}} (\bibinfo {year} {2026}),\ \Eprint {https://arxiv.org/abs/2606.06653} {arXiv:2606.06653} \BibitemShut {NoStop}%
\bibitem [{\citenamefont {Hammer}\ \emph {et~al.}(2026)\citenamefont {Hammer}, \citenamefont {Rylands},\ and\ \citenamefont {Carollo}}]{hammer26}%
  \BibitemOpen
  \bibfield  {author} {\bibinfo {author} {\bibfnamefont {L.}~\bibnamefont {Hammer}}, \bibinfo {author} {\bibfnamefont {C.}~\bibnamefont {Rylands}},\ and\ \bibinfo {author} {\bibfnamefont {F.}~\bibnamefont {Carollo}},\ }\href@noop {} {\bibinfo {title} {{Asymmetry dynamics and nonequilibrium symmetry-breaking phase transitions}}} (\bibinfo {year} {2026}),\ \Eprint {https://arxiv.org/abs/2606.07188} {arXiv:2606.07188} \BibitemShut {NoStop}%
\bibitem [{\citenamefont {Vescovo}\ \emph {et~al.}(2026)\citenamefont {Vescovo}, \citenamefont {Calabrese},\ and\ \citenamefont {Ares}}]{vescovo26}%
  \BibitemOpen
  \bibfield  {author} {\bibinfo {author} {\bibfnamefont {M.}~\bibnamefont {Vescovo}}, \bibinfo {author} {\bibfnamefont {P.}~\bibnamefont {Calabrese}},\ and\ \bibinfo {author} {\bibfnamefont {F.}~\bibnamefont {Ares}},\ }\href@noop {} {\bibinfo {title} {{Entanglement asymmetry and quantum Mpemba effect for Kramers-Wannier duality}}} (\bibinfo {year} {2026}),\ \Eprint {https://arxiv.org/abs/2607.21226} {arXiv:2607.21226} \BibitemShut {NoStop}%
\bibitem [{\citenamefont {Ares}\ \emph {et~al.}(2023{\natexlab{b}})\citenamefont {Ares}, \citenamefont {Murciano}, \citenamefont {Vernier},\ and\ \citenamefont {Calabrese}}]{Ares:2023kcz}%
  \BibitemOpen
  \bibfield  {author} {\bibinfo {author} {\bibfnamefont {F.}~\bibnamefont {Ares}}, \bibinfo {author} {\bibfnamefont {S.}~\bibnamefont {Murciano}}, \bibinfo {author} {\bibfnamefont {E.}~\bibnamefont {Vernier}},\ and\ \bibinfo {author} {\bibfnamefont {P.}~\bibnamefont {Calabrese}},\ }\bibfield  {title} {\bibinfo {title} {{Lack of symmetry restoration after a quantum quench: An entanglement asymmetry study}},\ }\href {https://doi.org/10.21468/SciPostPhys.15.3.089} {\bibfield  {journal} {\bibinfo  {journal} {SciPost Phys.}\ }\textbf {\bibinfo {volume} {15}},\ \bibinfo {pages} {089} (\bibinfo {year} {2023}{\natexlab{b}})}\BibitemShut {NoStop}%
\bibitem [{\citenamefont {Caceffo}\ \emph {et~al.}(2024)\citenamefont {Caceffo}, \citenamefont {Murciano},\ and\ \citenamefont {Alba}}]{Caceffo_2024}%
  \BibitemOpen
  \bibfield  {author} {\bibinfo {author} {\bibfnamefont {F.}~\bibnamefont {Caceffo}}, \bibinfo {author} {\bibfnamefont {S.}~\bibnamefont {Murciano}},\ and\ \bibinfo {author} {\bibfnamefont {V.}~\bibnamefont {Alba}},\ }\bibfield  {title} {\bibinfo {title} {{Entangled multiplets, asymmetry, and quantum Mpemba effect in dissipative systems}},\ }\href {https://doi.org/10.1088/1742-5468/ad4537} {\bibfield  {journal} {\bibinfo  {journal} {J. Stat. Mech.}\ ,\ \bibinfo {pages} {063103}} (\bibinfo {year} {2024})}\BibitemShut {NoStop}%
\bibitem [{\citenamefont {Yamashika}\ \emph {et~al.}(2025)\citenamefont {Yamashika}, \citenamefont {Calabrese},\ and\ \citenamefont {Ares}}]{yamashika25}%
  \BibitemOpen
  \bibfield  {author} {\bibinfo {author} {\bibfnamefont {S.}~\bibnamefont {Yamashika}}, \bibinfo {author} {\bibfnamefont {P.}~\bibnamefont {Calabrese}},\ and\ \bibinfo {author} {\bibfnamefont {F.}~\bibnamefont {Ares}},\ }\bibfield  {title} {\bibinfo {title} {{Quenching from superfluid to free bosons in two dimensions: Entanglement, symmetries, and the quantum Mpemba effect}},\ }\href {https://doi.org/10.1103/PhysRevA.111.043304} {\bibfield  {journal} {\bibinfo  {journal} {Phys. Rev. A}\ }\textbf {\bibinfo {volume} {111}},\ \bibinfo {pages} {043304} (\bibinfo {year} {2025})}\BibitemShut {NoStop}%
\bibitem [{\citenamefont {Hara}\ \emph {et~al.}(2026)\citenamefont {Hara}, \citenamefont {Endo},\ and\ \citenamefont {Yamashika}}]{hara26}%
  \BibitemOpen
  \bibfield  {author} {\bibinfo {author} {\bibfnamefont {R.}~\bibnamefont {Hara}}, \bibinfo {author} {\bibfnamefont {S.}~\bibnamefont {Endo}},\ and\ \bibinfo {author} {\bibfnamefont {S.}~\bibnamefont {Yamashika}},\ }\bibfield  {title} {\bibinfo {title} {Dynamics of entanglement asymmetry for space-inversion symmetry of free fermions on honeycomb lattices},\ }\href {https://doi.org/10.1103/lpz6-3v48} {\bibfield  {journal} {\bibinfo  {journal} {Phys. Rev. B}\ }\textbf {\bibinfo {volume} {113}},\ \bibinfo {pages} {144313} (\bibinfo {year} {2026})}\BibitemShut {NoStop}%
\bibitem [{\citenamefont {Giulio}\ \emph {et~al.}(2025)\citenamefont {Giulio}, \citenamefont {Turkeshi},\ and\ \citenamefont {Murciano}}]{digiulio25}%
  \BibitemOpen
  \bibfield  {author} {\bibinfo {author} {\bibfnamefont {G.~D.}\ \bibnamefont {Giulio}}, \bibinfo {author} {\bibfnamefont {X.}~\bibnamefont {Turkeshi}},\ and\ \bibinfo {author} {\bibfnamefont {S.}~\bibnamefont {Murciano}},\ }\bibfield  {title} {\bibinfo {title} {Measurement-induced symmetry restoration and quantum mpemba effect},\ }\href {https://doi.org/10.3390/e27040407} {\bibfield  {journal} {\bibinfo  {journal} {Entropy}\ }\textbf {\bibinfo {volume} {27}},\ \bibinfo {pages} {407} (\bibinfo {year} {2025})}\BibitemShut {NoStop}%
\bibitem [{\citenamefont {Ares}\ \emph {et~al.}(2025{\natexlab{c}})\citenamefont {Ares}, \citenamefont {Calabrese},\ and\ \citenamefont {Murciano}}]{ares25nat}%
  \BibitemOpen
  \bibfield  {author} {\bibinfo {author} {\bibfnamefont {F.}~\bibnamefont {Ares}}, \bibinfo {author} {\bibfnamefont {P.}~\bibnamefont {Calabrese}},\ and\ \bibinfo {author} {\bibfnamefont {S.}~\bibnamefont {Murciano}},\ }\bibfield  {title} {\bibinfo {title} {{The quantum Mpemba effects}},\ }\href {https://doi.org/10.1038/s42254-025-00838-0} {\bibfield  {journal} {\bibinfo  {journal} {Nat. Rev. Phys.}\ }\textbf {\bibinfo {volume} {7}},\ \bibinfo {pages} {451} (\bibinfo {year} {2025}{\natexlab{c}})}\BibitemShut {NoStop}%
\bibitem [{\citenamefont {Teza}\ \emph {et~al.}(2026)\citenamefont {Teza}, \citenamefont {Bechhoefer}, \citenamefont {Lasanta}, \citenamefont {Raz},\ and\ \citenamefont {Vucelja}}]{teza26}%
  \BibitemOpen
  \bibfield  {author} {\bibinfo {author} {\bibfnamefont {G.}~\bibnamefont {Teza}}, \bibinfo {author} {\bibfnamefont {J.}~\bibnamefont {Bechhoefer}}, \bibinfo {author} {\bibfnamefont {A.}~\bibnamefont {Lasanta}}, \bibinfo {author} {\bibfnamefont {O.}~\bibnamefont {Raz}},\ and\ \bibinfo {author} {\bibfnamefont {M.}~\bibnamefont {Vucelja}},\ }\bibfield  {title} {\bibinfo {title} {{Speedups in nonequilibrium thermal relaxation: Mpemba and related effects}},\ }\href {https://doi.org/10.1016/j.physrep.2025.10.009} {\bibfield  {journal} {\bibinfo  {journal} {Phys. Rep.}\ }\textbf {\bibinfo {volume} {1164}},\ \bibinfo {pages} {1} (\bibinfo {year} {2026})}\BibitemShut {NoStop}%
\bibitem [{\citenamefont {Yu}\ \emph {et~al.}(2025)\citenamefont {Yu}, \citenamefont {Liu},\ and\ \citenamefont {Zhang}}]{yu2025}%
  \BibitemOpen
  \bibfield  {author} {\bibinfo {author} {\bibfnamefont {H.}~\bibnamefont {Yu}}, \bibinfo {author} {\bibfnamefont {S.}~\bibnamefont {Liu}},\ and\ \bibinfo {author} {\bibfnamefont {S.-X.}\ \bibnamefont {Zhang}},\ }\bibfield  {title} {\bibinfo {title} {{Quantum Mpemba effects from symmetry perspectives}},\ }\href {https://link.springer.com/content/pdf/10.1007/s43673-025-00157-7.pdf} {\bibfield  {journal} {\bibinfo  {journal} {AAPPS Bulletin}\ }\textbf {\bibinfo {volume} {35}} (\bibinfo {year} {2025})}\BibitemShut {NoStop}%
\bibitem [{\citenamefont {Calabrese}(2026)}]{calabrese26}%
  \BibitemOpen
  \bibfield  {author} {\bibinfo {author} {\bibfnamefont {P.}~\bibnamefont {Calabrese}},\ }\bibfield  {title} {\bibinfo {title} {{The quantum Mpemba effect in closed systems: from theory to experiment}},\ }\href {https://iopscience.iop.org/article/10.1088/1742-5468/ae4bb6/meta} {\bibfield  {journal} {\bibinfo  {journal} {J. Stat. Mech.}\ ,\ \bibinfo {pages} {034002}} (\bibinfo {year} {2026})}\BibitemShut {NoStop}%
\bibitem [{\citenamefont {Chalas}\ \emph {et~al.}(2024)\citenamefont {Chalas}, \citenamefont {Ares}, \citenamefont {Rylands},\ and\ \citenamefont {Calabrese}}]{Chalas_2024}%
  \BibitemOpen
  \bibfield  {author} {\bibinfo {author} {\bibfnamefont {K.}~\bibnamefont {Chalas}}, \bibinfo {author} {\bibfnamefont {F.}~\bibnamefont {Ares}}, \bibinfo {author} {\bibfnamefont {C.}~\bibnamefont {Rylands}},\ and\ \bibinfo {author} {\bibfnamefont {P.}~\bibnamefont {Calabrese}},\ }\bibfield  {title} {\bibinfo {title} {{Multiple crossings during dynamical symmetry restoration and implications for the quantum Mpemba effect}},\ }\href {https://doi.org/10.1088/1742-5468/ad769c} {\bibfield  {journal} {\bibinfo  {journal} {J. Stat. Mech.}\ ,\ \bibinfo {pages} {103101}} (\bibinfo {year} {2024})}\BibitemShut {NoStop}%
\bibitem [{\citenamefont {Chatterjee}\ \emph {et~al.}(2024)\citenamefont {Chatterjee}, \citenamefont {Takada},\ and\ \citenamefont {Hayakawa}}]{chatterjee24}%
  \BibitemOpen
  \bibfield  {author} {\bibinfo {author} {\bibfnamefont {A.~K.}\ \bibnamefont {Chatterjee}}, \bibinfo {author} {\bibfnamefont {S.}~\bibnamefont {Takada}},\ and\ \bibinfo {author} {\bibfnamefont {H.}~\bibnamefont {Hayakawa}},\ }\bibfield  {title} {\bibinfo {title} {{Multiple quantum Mpemba effect: exceptional points and oscillations}},\ }\href {https://doi.org/10.1103/PhysRevA.110.022213} {\bibfield  {journal} {\bibinfo  {journal} {Phys. Rev. A}\ }\textbf {\bibinfo {volume} {110}},\ \bibinfo {pages} {022213} (\bibinfo {year} {2024})}\BibitemShut {NoStop}%
\bibitem [{\citenamefont {McRoberts}(2026)}]{mcroberts26}%
  \BibitemOpen
  \bibfield  {author} {\bibinfo {author} {\bibfnamefont {A.~J.}\ \bibnamefont {McRoberts}},\ }\bibfield  {title} {\bibinfo {title} {{Integrability-breaking-induced Mpemba effect in spin chains}},\ }\href {https://doi.org/10.1088/1751-8121/ae7920} {\bibfield  {journal} {\bibinfo  {journal} {J. Phys. A: Math. Theor.}\ }\textbf {\bibinfo {volume} {59}},\ \bibinfo {pages} {24LT01} (\bibinfo {year} {2026})}\BibitemShut {NoStop}%
\bibitem [{\citenamefont {Fan}\ \emph {et~al.}(2020)\citenamefont {Fan}, \citenamefont {Gu}, \citenamefont {Vishwanath},\ and\ \citenamefont {Wen}}]{Fan_2020}%
  \BibitemOpen
  \bibfield  {author} {\bibinfo {author} {\bibfnamefont {R.}~\bibnamefont {Fan}}, \bibinfo {author} {\bibfnamefont {Y.}~\bibnamefont {Gu}}, \bibinfo {author} {\bibfnamefont {A.}~\bibnamefont {Vishwanath}},\ and\ \bibinfo {author} {\bibfnamefont {X.}~\bibnamefont {Wen}},\ }\bibfield  {title} {\bibinfo {title} {{Emergent Spatial Structure and Entanglement Localization in Floquet Conformal Field Theory}},\ }\href {https://doi.org/10.1103/physrevx.10.031036} {\bibfield  {journal} {\bibinfo  {journal} {Physical Review X}\ }\textbf {\bibinfo {volume} {10}},\ \bibinfo {pages} {031036} (\bibinfo {year} {2020})}\BibitemShut {NoStop}%
\bibitem [{\citenamefont {Calabrese}\ and\ \citenamefont {Cardy}(2004)}]{cc04}%
  \BibitemOpen
  \bibfield  {author} {\bibinfo {author} {\bibfnamefont {P.}~\bibnamefont {Calabrese}}\ and\ \bibinfo {author} {\bibfnamefont {J.}~\bibnamefont {Cardy}},\ }\bibfield  {title} {\bibinfo {title} {Entanglement entropy and quantum field theory},\ }\href {https://doi.org/10.1088/1742-5468/2004/06/P06002} {\bibfield  {journal} {\bibinfo  {journal} {J. Stat. Mech.}\ ,\ \bibinfo {pages} {P06002}} (\bibinfo {year} {2004})}\BibitemShut {NoStop}%
\bibitem [{\citenamefont {Calabrese}\ and\ \citenamefont {Cardy}(2009)}]{cc09}%
  \BibitemOpen
  \bibfield  {author} {\bibinfo {author} {\bibfnamefont {P.}~\bibnamefont {Calabrese}}\ and\ \bibinfo {author} {\bibfnamefont {J.}~\bibnamefont {Cardy}},\ }\bibfield  {title} {\bibinfo {title} {Entanglement entropy and conformal field theory},\ }\href {https://doi.org/10.1088/1751-8113/42/50/504005} {\bibfield  {journal} {\bibinfo  {journal} {J. Phys. A: Math. Theor.}\ }\textbf {\bibinfo {volume} {42}},\ \bibinfo {pages} {504005} (\bibinfo {year} {2009})}\BibitemShut {NoStop}%
\bibitem [{\citenamefont {Fossati}\ \emph {et~al.}(2024)\citenamefont {Fossati}, \citenamefont {Ares}, \citenamefont {Dubail},\ and\ \citenamefont {Calabrese}}]{fossati24}%
  \BibitemOpen
  \bibfield  {author} {\bibinfo {author} {\bibfnamefont {M.}~\bibnamefont {Fossati}}, \bibinfo {author} {\bibfnamefont {F.}~\bibnamefont {Ares}}, \bibinfo {author} {\bibfnamefont {J.}~\bibnamefont {Dubail}},\ and\ \bibinfo {author} {\bibfnamefont {P.}~\bibnamefont {Calabrese}},\ }\bibfield  {title} {\bibinfo {title} {{Entanglement asymmetry in CFT and its relation to non-topological defects}},\ }\href {https://doi.org/https://doi.org/10.1007/JHEP05(2024)059} {\bibfield  {journal} {\bibinfo  {journal} {{J. High Energy Phys.}}\ }\textbf {\bibinfo {volume} {05}},\ \bibinfo {pages} {059} (\bibinfo {year} {2024})}\BibitemShut {NoStop}%
\bibitem [{\citenamefont {Chen}\ and\ \citenamefont {Chen}(2024)}]{chenchen24}%
  \BibitemOpen
  \bibfield  {author} {\bibinfo {author} {\bibfnamefont {M.}~\bibnamefont {Chen}}\ and\ \bibinfo {author} {\bibfnamefont {H.-H.}\ \bibnamefont {Chen}},\ }\bibfield  {title} {\bibinfo {title} {Rényi entanglement asymmetry in (1+1)-dimensional conformal field theories},\ }\href {https://doi.org/10.1103/PhysRevD.109.065009} {\bibfield  {journal} {\bibinfo  {journal} {Phys. Rev. D}\ }\textbf {\bibinfo {volume} {109}},\ \bibinfo {pages} {065009} (\bibinfo {year} {2024})}\BibitemShut {NoStop}%
\bibitem [{\citenamefont {Kusuki}\ \emph {et~al.}(2025)\citenamefont {Kusuki}, \citenamefont {Murciano}, \citenamefont {Ooguri},\ and\ \citenamefont {Pal}}]{kusuki25}%
  \BibitemOpen
  \bibfield  {author} {\bibinfo {author} {\bibfnamefont {Y.}~\bibnamefont {Kusuki}}, \bibinfo {author} {\bibfnamefont {S.}~\bibnamefont {Murciano}}, \bibinfo {author} {\bibfnamefont {H.}~\bibnamefont {Ooguri}},\ and\ \bibinfo {author} {\bibfnamefont {S.}~\bibnamefont {Pal}},\ }\bibfield  {title} {\bibinfo {title} {Entanglement asymmetry and symmetry defects in boundary conformal field theory},\ }\href {https://doi.org/10.1007/JHEP01(2025)057} {\bibfield  {journal} {\bibinfo  {journal} {{J. High Energy Phys.}}\ }\textbf {\bibinfo {volume} {01}},\ \bibinfo {pages} {057} (\bibinfo {year} {2025})}\BibitemShut {NoStop}%
\bibitem [{\citenamefont {Fossati}\ \emph {et~al.}(2025)\citenamefont {Fossati}, \citenamefont {Rylands},\ and\ \citenamefont {Calabrese}}]{fossati25}%
  \BibitemOpen
  \bibfield  {author} {\bibinfo {author} {\bibfnamefont {M.}~\bibnamefont {Fossati}}, \bibinfo {author} {\bibfnamefont {C.}~\bibnamefont {Rylands}},\ and\ \bibinfo {author} {\bibfnamefont {P.}~\bibnamefont {Calabrese}},\ }\bibfield  {title} {\bibinfo {title} {{Entanglement asymmetry in CFT with boundary symmetry breaking}},\ }\href {https://doi.org/10.1007/JHEP06(2025)089} {\bibfield  {journal} {\bibinfo  {journal} {{J. High Energy Phys.}}\ }\textbf {\bibinfo {volume} {06}},\ \bibinfo {pages} {089} (\bibinfo {year} {2025})}\BibitemShut {NoStop}%
\bibitem [{\citenamefont {Lastres}\ \emph {et~al.}(2025)\citenamefont {Lastres}, \citenamefont {Murciano}, \citenamefont {Ares},\ and\ \citenamefont {Calabrese}}]{lastres25}%
  \BibitemOpen
  \bibfield  {author} {\bibinfo {author} {\bibfnamefont {M.}~\bibnamefont {Lastres}}, \bibinfo {author} {\bibfnamefont {S.}~\bibnamefont {Murciano}}, \bibinfo {author} {\bibfnamefont {F.}~\bibnamefont {Ares}},\ and\ \bibinfo {author} {\bibfnamefont {P.}~\bibnamefont {Calabrese}},\ }\bibfield  {title} {\bibinfo {title} {{Entanglement asymmetry in the critical XXZ spin chain}},\ }\href {https://doi.org/10.1088/1742-5468/ada497} {\bibfield  {journal} {\bibinfo  {journal} {J. Stat. Mech.}\ ,\ \bibinfo {pages} {013107}} (\bibinfo {year} {2025})}\BibitemShut {NoStop}%
\bibitem [{\citenamefont {Benini}\ \emph {et~al.}(2025{\natexlab{b}})\citenamefont {Benini}, \citenamefont {Calabrese}, \citenamefont {Fossati}, \citenamefont {Singh},\ and\ \citenamefont {Venuti}}]{venuti25}%
  \BibitemOpen
  \bibfield  {author} {\bibinfo {author} {\bibfnamefont {F.}~\bibnamefont {Benini}}, \bibinfo {author} {\bibfnamefont {P.}~\bibnamefont {Calabrese}}, \bibinfo {author} {\bibfnamefont {M.}~\bibnamefont {Fossati}}, \bibinfo {author} {\bibfnamefont {A.~H.}\ \bibnamefont {Singh}},\ and\ \bibinfo {author} {\bibfnamefont {M.}~\bibnamefont {Venuti}},\ }\href@noop {} {\bibinfo {title} {{Entanglement asymmetry for higher and noninvertible symmetries}}} (\bibinfo {year} {2025}{\natexlab{b}}),\ \Eprint {https://arxiv.org/abs/2509.16311} {arXiv:2509.16311} \BibitemShut {NoStop}%
\bibitem [{\citenamefont {Benini}\ \emph {et~al.}(2025{\natexlab{c}})\citenamefont {Benini}, \citenamefont {Garcia-Valdecasas},\ and\ \citenamefont {Vitouladitis}}]{valdecasas25}%
  \BibitemOpen
  \bibfield  {author} {\bibinfo {author} {\bibfnamefont {F.}~\bibnamefont {Benini}}, \bibinfo {author} {\bibfnamefont {E.}~\bibnamefont {Garcia-Valdecasas}},\ and\ \bibinfo {author} {\bibfnamefont {S.}~\bibnamefont {Vitouladitis}},\ }\href@noop {} {\bibinfo {title} {{Higher-form entanglement asymmetry. Part I. The limits of symmetry breaking}}} (\bibinfo {year} {2025}{\natexlab{c}}),\ \Eprint {https://arxiv.org/abs/2512.15898} {2512.15898} \BibitemShut {NoStop}%
\bibitem [{\citenamefont {Fossati}\ \emph {et~al.}(2026)\citenamefont {Fossati}, \citenamefont {Rylands}, \citenamefont {Grosfeld}, \citenamefont {Sela},\ and\ \citenamefont {Calabrese}}]{fossati26}%
  \BibitemOpen
  \bibfield  {author} {\bibinfo {author} {\bibfnamefont {M.}~\bibnamefont {Fossati}}, \bibinfo {author} {\bibfnamefont {C.}~\bibnamefont {Rylands}}, \bibinfo {author} {\bibfnamefont {E.}~\bibnamefont {Grosfeld}}, \bibinfo {author} {\bibfnamefont {E.}~\bibnamefont {Sela}},\ and\ \bibinfo {author} {\bibfnamefont {P.}~\bibnamefont {Calabrese}},\ }\href@noop {} {\bibinfo {title} {{Boundary quenches in (1+ 1)-dimensional conformal field theory}}} (\bibinfo {year} {2026}),\ \Eprint {https://arxiv.org/abs/2607.19166} {arXiv:2607.19166} \BibitemShut {NoStop}%
\bibitem [{\citenamefont {Lu}\ and\ \citenamefont {Raz}(2017)}]{lu17}%
  \BibitemOpen
  \bibfield  {author} {\bibinfo {author} {\bibfnamefont {Z.}~\bibnamefont {Lu}}\ and\ \bibinfo {author} {\bibfnamefont {O.}~\bibnamefont {Raz}},\ }\bibfield  {title} {\bibinfo {title} {{Nonequilibrium thermodynamics of the Markovian Mpemba effect and its inverse}},\ }\href {https://doi.org/10.1073/pnas.1701264114} {\bibfield  {journal} {\bibinfo  {journal} {PNAS}\ }\textbf {\bibinfo {volume} {114}},\ \bibinfo {pages} {5083} (\bibinfo {year} {2017})}\BibitemShut {NoStop}%
\bibitem [{\citenamefont {Kumar}\ \emph {et~al.}(2022)\citenamefont {Kumar}, \citenamefont {Chetrite},\ and\ \citenamefont {Bechhoefer}}]{kumar22}%
  \BibitemOpen
  \bibfield  {author} {\bibinfo {author} {\bibfnamefont {A.}~\bibnamefont {Kumar}}, \bibinfo {author} {\bibfnamefont {R.}~\bibnamefont {Chetrite}},\ and\ \bibinfo {author} {\bibfnamefont {J.}~\bibnamefont {Bechhoefer}},\ }\bibfield  {title} {\bibinfo {title} {{Anomalous heating in a colloidal system}},\ }\href {https://doi.org/10.1073/pnas.2118484119} {\bibfield  {journal} {\bibinfo  {journal} {PNAS}\ }\textbf {\bibinfo {volume} {119}},\ \bibinfo {pages} {e2118484119} (\bibinfo {year} {2022})}\BibitemShut {NoStop}%
\bibitem [{\citenamefont {Aharony~Shapira}\ \emph {et~al.}(2024)\citenamefont {Aharony~Shapira}, \citenamefont {Shapira}, \citenamefont {Markov}, \citenamefont {Teza}, \citenamefont {Akerman}, \citenamefont {Raz},\ and\ \citenamefont {Ozeri}}]{AharonyShapira:2024nrt}%
  \BibitemOpen
  \bibfield  {author} {\bibinfo {author} {\bibfnamefont {S.}~\bibnamefont {Aharony~Shapira}}, \bibinfo {author} {\bibfnamefont {Y.}~\bibnamefont {Shapira}}, \bibinfo {author} {\bibfnamefont {J.}~\bibnamefont {Markov}}, \bibinfo {author} {\bibfnamefont {G.}~\bibnamefont {Teza}}, \bibinfo {author} {\bibfnamefont {N.}~\bibnamefont {Akerman}}, \bibinfo {author} {\bibfnamefont {O.}~\bibnamefont {Raz}},\ and\ \bibinfo {author} {\bibfnamefont {R.}~\bibnamefont {Ozeri}},\ }\bibfield  {title} {\bibinfo {title} {{Inverse Mpemba Effect Demonstrated on a Single Trapped Ion Qubit}},\ }\href {https://doi.org/10.1103/PhysRevLett.133.010403} {\bibfield  {journal} {\bibinfo  {journal} {Phys. Rev. Lett.}\ }\textbf {\bibinfo {volume} {133}},\ \bibinfo {pages} {010403} (\bibinfo {year} {2024})}\BibitemShut {NoStop}%
\bibitem [{\citenamefont {Sugimoto}\ \emph {et~al.}(2025)\citenamefont {Sugimoto}, \citenamefont {Kuwahara},\ and\ \citenamefont {Saito}}]{Sugimoto:2025pgk}%
  \BibitemOpen
  \bibfield  {author} {\bibinfo {author} {\bibfnamefont {K.}~\bibnamefont {Sugimoto}}, \bibinfo {author} {\bibfnamefont {T.}~\bibnamefont {Kuwahara}},\ and\ \bibinfo {author} {\bibfnamefont {K.}~\bibnamefont {Saito}},\ }\href@noop {} {\bibinfo {title} {{Prethermal inverse Mpemba effect}}} (\bibinfo {year} {2025}),\ \Eprint {https://arxiv.org/abs/2507.04669} {arXiv:2507.04669} \BibitemShut {NoStop}%
\bibitem [{\citenamefont {Ares}\ \emph {et~al.}(2026)\citenamefont {Ares}, \citenamefont {Das},\ and\ \citenamefont {Kundu}}]{Das:toappear}%
  \BibitemOpen
  \bibfield  {author} {\bibinfo {author} {\bibfnamefont {F.}~\bibnamefont {Ares}}, \bibinfo {author} {\bibfnamefont {J.}~\bibnamefont {Das}},\ and\ \bibinfo {author} {\bibfnamefont {A.}~\bibnamefont {Kundu}},\ }\href@noop {} {\bibinfo {title} {{Entanglement Asymmetry and Quantum Mpemba Effect in Driven Two-dimensional CFTs}}} (\bibinfo {year} {2026}),\ \bibinfo {note} {to appear}\BibitemShut {NoStop}%
\end{thebibliography}%

\appendix

\section{End Matter} 
\label{endmat} 
\raggedbottom

Here, we provide details that are essential for deriving the results presented in the main text, as well as some additional results.

We consider a two-dimensional CFT on the cylinder with a global
U$(1)$ symmetry generated by the conserved charge $Q$. Given a density matrix
$\rho$, and assuming a standard factorization of the Hilbert
space, we obtain the reduced density matrix $\rho_A$ associated
with a spatial bipartition $A\cup \bar{A}$. We further assume that the
global charge admits an additive decomposition,
$Q=Q_A\otimes{\mathbbm 1}_{\bar{A}}+{\mathbbm 1}_A\otimes Q_{\bar{A}}$.
The reduced density matrix $\rho_A$ may contain off-diagonal
elements in the eigenbasis of $Q_A$, resulting in a non-vanishing
commutator $[\rho_A,Q_A]\neq0$. In this case, the reduced state
is not invariant under the U$(1)$ symmetry. Such symmetry
breaking can be realized by a local operator that does not
commute with the charge acting, for example, on the vacuum,
thereby creating a symmetry-breaking excited state~\cite{Benini:2024xjv}. Here, we
consider a spinless primary operator $V$ of scaling dimension
$\Delta$ and definite U$(1)$ charge $q$, \textit{i.e.},
$e^{i\alpha Q}V e^{-i\alpha Q}=e^{iq\alpha}V$.
We define the corresponding local operator as
\begin{equation}
    O(w)=e^{i\kappa V(w)}
    =
    1+i\kappa V(w)+\mathcal O(\kappa^2),
    \label{eq:coherent-operator}
\end{equation}
where $\kappa$ is a small perturbative parameter. In our Floquet protocol, the initial excited state is prepared by inserting $O$ at $w_-=-\tau_0+i\eta$, $|\psi_0\rangle=O(w_-)|0\rangle$, with $\tau_0>0$. Using the Heisenberg picture, where the operator $O$ evolves under the drive, the state of the system after $N$ Floquet cycles is 
\begin{equation}
|\psi_N\rangle=O_N(w_-)|0\rangle=[1+i\kappa V_N(w_-)+\mathcal O(\kappa^2)]|0\rangle,
\end{equation}
where $O_N(w_-)\equiv U_N^\dagger O(w_-)U_N$ and $V_N(w_-)\equiv U_N^\dagger V(w_-)U_N$.
Since the evolution is generated by the CFT stress tensor, $U_N$ commutes with the global U$(1)$ charge. Hence, $V_N$ carries the same U$(1)$ charge as $V$. Moreover, since $V$ is primary and the evolution acts by a conformal transformation, its scaling dimension remains $\Delta$.

To obtain the entanglement asymmetry~\eqref{eq:def_EA}, we consider 
the R\'enyi-$n$ entanglement asymmetry~\cite{ares23asymmetry}:
\begin{equation}
\Delta S_{n, A}=\frac{1}{1-n}\log\frac{{\rm Tr}(\rho_{A, Q}^n)}{{\rm Tr}(\rho_A^n)}.
\end{equation}
In the limit $n\to 1$, Eq.~\eqref{eq:def_EA} is recovered. The symmetrized 
density matrix reads
\begin{equation}
\rho_{A, Q}=\int_{-\pi}^\pi\frac{d\alpha}{2\pi}e^{i\alpha Q_A}\rho_A e^{-i\alpha Q_A}.
\end{equation}
Therefore,
\begin{equation}
\begin{aligned}
    {\rm Tr}(\rho_{A,Q}^n)
    &=
    \int_{-\pi}^{\pi}
    \frac{d\gamma_1\cdots d\gamma_{n-1}}{(2\pi)^{n-1}}\,
    \\
    &\quad\times
    {\rm Tr}\left[
    \rho_A e^{i\gamma_1 Q_A}
    \cdots
    \rho_A e^{i\gamma_{n}Q_A}
    \right],
\end{aligned}
    \label{eq:rhoQn-charged-moments}
\end{equation}
with $\gamma_1+\cdots+\gamma_n=0$.
For a two-dimensional CFT in a state of the form $|\psi_N\rangle$, the path integral representation of the trace in Eq.~\eqref{eq:rhoQn-charged-moments}, with insertions $e^{i\gamma_j Q_A}$, has been evaluated perturbatively in the limit $\kappa\ll1$ in Ref.~\cite{Benini:2024xjv}, where the calculation is discussed in detail. Applying these results directly to our case, we obtain at leading order in $\kappa$:
\begin{equation}
\Delta S_{n,A}(N) = \frac{\kappa^2}{n-1}
\sum_{\substack{j,k=1\\j\ne k}}^{n}
\left\langle V_{N,j}(w_-) V_{N,k}^\dagger(w_+) \right\rangle_{{\cal G}} + \mathcal{O}(\kappa^4).
\label{repcylinder}
\end{equation}
Here, $w_+=-\bar w_-=\tau_0+i\eta$ and $\langle \bullet\rangle_\mathcal{G}$ denote the CFT correlator on the replicated Riemann surface $\mathcal{G}$, consisting in our case of $n$ cylinders of circumference $L$ cyclically glued along the branch cut that defines the subsystem $A$. The operator $V_{N,j}$ denotes the $N$-evolved operator $V$ on the $j$-th replica.  

We can now apply Eq.~\eqref{eq:primary-time-evolution} in Eq.~\eqref{repcylinder}. It then remains to evaluate $\left\langle V_j(w_{-,N})V_k^\dagger(w_{+,N})\right\rangle_{\mathcal G}$.
To this end, we map the replica geometry to a convenient coordinate system. Given a subsystem with endpoints at $\ell_{A_1}$ and at $\ell_{A_2}$, the branch cut of $\mathcal{G}$ can then be conveniently mapped to the semi-infinite positive real half line using a further M\"{o}bius transformation,
\begin{eqnarray}\label{eq:rindler}
    Z =\Phi_A(u)=e^{i\delta} \frac{u-u_1}{u_2 - u} \ , \quad u_{1,2} = e^{2\pi i \ell_{A_{1,2}}/L} \ ,
 \end{eqnarray}
such that the branch points are mapped to $0$ and $\infty$. Here the phase \(\delta\) is chosen such that the image of the subsystem \(A\) lies on the positive real axis. As a result, the replicated cylinder correlator in (\ref{repcylinder}) is eventually expressed in terms of the two-point functions of $V$ on a replicated Rindler geometry. The $n$-sheeted Rindler geometry is subsequently uniformized by the covering map $\xi = Z^{1/n}$~\cite{Benini:2024xjv, Das:toappear}, reducing the correlators in Eq.~\eqref{repcylinder} to two-point functions of $V$ on the complex plane.

It is useful to summarize the sequence of coordinate transformations entering the calculation:
\begin{equation*}
\begin{tikzcd}[
    column sep=3.2em,
    row sep=2.0em,
    cells={nodes={
        draw=chartedge,
        fill=chartfill,
        rounded corners=2pt,
        line width=0.45pt,
        inner xsep=3pt,
        inner ysep=3pt
    }}
]
{\text{\bfseries Rep. cylinder }(w)}
\arrow[r, "u=e^{2\pi w/L}"]
&
{\text{\bfseries Rep. plane }(u)}
\arrow[d, "f^N"']
\\
&
{
\begin{array}{c}
\text{\bfseries Floquet-evolved}\\[-1mm]
\text{\bfseries rep. plane }(u_N)
\end{array}}
\arrow[d, "Z=e^{i\delta}\frac{u-u_1}{u_2-u}"']
\\
{\text{\bfseries Plane }(\xi_N)}
&
{
\begin{array}{c}
\text{\bfseries Floquet-evolved}\\[-0.7mm]
\text{\bfseries rep. Rindler }(Z_N)
\end{array}}
\arrow[l, "\xi=Z^{1/n}"']
\end{tikzcd}
\label{eq:coordinate-diagram}
\end{equation*}
Thus, the Floquet evolution is performed on the replicated complex plane before the evolved insertion point is mapped to the Rindler geometry used to evaluate the replicated correlation functions.

Following this sequence of conformal transformations and taking the replica limit $n\to1$, the leading contribution in $\kappa$ admits a closed-form expression for a primary operator $V$ of arbitrary scaling dimension $\Delta$. Writing
\begin{equation}
    \Delta S_A(N)
    =
    \kappa^2\Delta S_A^{(2)}(N)
    +\mathcal{O}(\kappa^4),
\end{equation}
we obtain
\begin{equation}
\begin{aligned}
    \Delta S_A^{(2)}(N)
    =\left|\frac{dZ_N(w)}{dw}
\right|_{w=w_-}^{2\Delta}
\frac{1}{r_N^{2\Delta}}\,
I_\Delta\!\left(
\tan\frac{\tau_N}{2}
\right).
\end{aligned}
\label{eq:anyDelta_EA}
\end{equation}
\clearpage
\onecolumngrid

\noindent
\parbox[t]{0.485\textwidth}{%
\vspace{0pt}

Here, $Z_N(w)=\Phi_A\!\left[f^N\!\left(e^{2\pi w/L}\right)\right]$ denotes the complete map from the cylinder to the Floquet-evolved Rindler coordinate. The conformal transformation of each primary operator produces one Jacobian contribution. Since the two operator insertions are related by complex conjugation, these contributions have equal magnitude. Their product therefore gives the Jacobian factor appearing in Eq.~\eqref{eq:anyDelta_EA}. We parametrize the transformed insertion points as $Z_{N,-}=r_Ne^{i(\pi-\tau_N)}$ and
$Z_{N,+}=Z_{N,-}^{*}=r_Ne^{-i(\pi-\tau_N)}$, with
$r_N>0$ and $0<\tau_N<\pi$. The function \(I_\Delta\) is a universal scaling function obtained after performing the sum over the replica-sheet two-point functions in Eq.~\eqref{repcylinder} and taking the analytic continuation to the replica limit \(n\to1\), following Ref.~\cite{Benini:2024xjv}. It is given by
\begin{equation}
\begin{aligned}
I_\Delta(x)
={}&
\frac{\sqrt{\pi}\,\Gamma(\Delta+1)}
     {4^\Delta\Gamma\!\left(\Delta+\frac12\right)}
\Bigg[
1-(1-x^2)
\frac{\Delta}{\Delta+\frac12}
\\
&\qquad\times
{}_2F_1\!\left(
1,\frac12-\Delta;
\frac32+\Delta;
-x^2
\right)
\Bigg].
\end{aligned}
\label{eq:I_Delta}
\end{equation}
Thus, the dependence on the insertion position, the subsystem
endpoints, and the Floquet evolution is encoded in the driven Rindler variables \(r_N\) and \(\tau_N\). See Ref.~\cite{Das:toappear} for further details.

In the main text, we illustrated crossings of the
entanglement-asymmetry curves in the heating phase and at the phase transition. In these two regimes, the fixed points on the unit circle determine a definite asymptotic flow of the operator insertion. Depending on the position of the relevant fixed point relative to the subsystem, the entanglement asymmetry can decay, grow, or approach a finite value. The crossings therefore occur on top of an underlying asymptotic tendency towards symmetry restoration, amplification of symmetry breaking, or saturation. It is natural to ask whether an exchange of the asymmetry ordering can also occur in the non-heating phase, where there is no fixed point on the unit circle in the complex plane and hence no relaxation towards an asymptotic configuration.
}%
\hfill
\parbox[t]{0.485\textwidth}{%
\vspace{0pt}

{%
\centering
\includegraphics[width=\linewidth]
{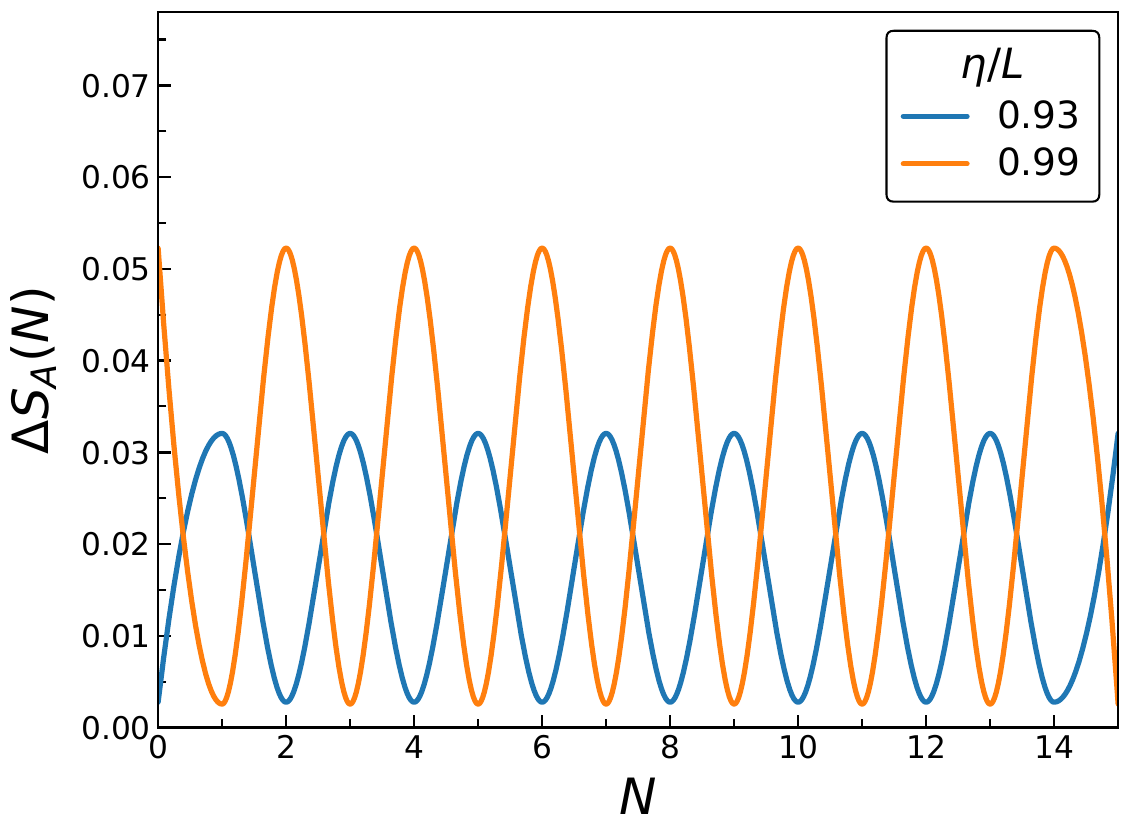}

\captionof{figure}{Evolution of the entanglement asymmetry
\(\Delta S_A(N)\) for two initial insertion positions \(\eta_1\) and \(\eta_2\) in the non-heating phase. We set \(\Delta=1\), \(\theta=0.15\), and \(\tau_0/L=0.02\), with \(A=[0,0.35L]\). The driving periods are chosen as
\(\left(T_0/L,T_1/L\right) =\left(1,\cosh(2\theta)/2\right)\).
The Floquet dynamics produces persistent crossings of the two
asymmetry curves.}
\label{fig:nonheating-mpemba}
\par
}
\vspace{2\baselineskip}

{\setlength{\baselineskip}{1.05\baselineskip}%
The answer is affirmative, as illustrated in Fig.~\ref{fig:nonheating-mpemba}. In this phase, the state-preparing insertion remains on a recurrent orbit, and the entanglement asymmetry displays persistent oscillations. Two states prepared by inserting the same charged operator at different positions can consequently exchange their ordering repeatedly: the state with the larger initial asymmetry becomes less asymmetric than the other, and vice versa, over successive drive cycles. Thus, the non-heating phase
supports repeated crossings even in the absence of relaxation.
Unlike the heating and phase-transition cases, these crossings do not reflect different relaxation rates towards a common late-time state; rather, they arise from the recurrent oscillations of the degree of local symmetry breaking generated by the elliptic Floquet dynamics.
\par
}
}

\end{document}